\documentclass[traditabstract]{aa}
\usepackage[OT1]{fontenc}
\usepackage[latin9]{inputenc}
\usepackage{booktabs}
\usepackage{textcomp}
\usepackage{natbib}
\usepackage[colorlinks=true,citecolor=blue]{hyperref}
\usepackage{color}
\usepackage{amsmath}
\usepackage{txfonts}
\usepackage{graphicx}
\usepackage{amssymb}
\usepackage{subfig}

\definecolor{darkgreen}{RGB}{0,208,20}
\definecolor{darkorange}{RGB}{222,100,0}

\def\d{\mathrm{d}}
\bibpunct{(}{)}{;}{a}{}{,}

\begin{document}

\defcitealias{Planck2013_XX}{PXX}
\defcitealias{Sheth2001}{SMT}

\title{X-ray galaxy clusters abundance and mass temperature scaling}
\titlerunning{X-ray galaxy clusters abundance and mass-temperature scaling}
\author{St\'ephane Ili\'c\inst{1,2}, Alain Blanchard\inst{1,2} \and Marian Douspis\inst{3,4}}
\authorrunning{S. Ilic\inst{1,2}, A. Blanchard\inst{1,2}  \& M. Douspis\inst{3,4} }
\institute{$^1$ Universit\'e de Toulouse; UPS-OMP; IRAP; Toulouse, France\\
 $^2$ CNRS; IRAP; 14, avenue Edouard Belin, F-31400 Toulouse, France\\
$^3$Institut d'Astrophysique Spatiale, Universit\'e Paris-Sud , UMR8617, Orsay, F-91405 \\
$^4$CNRS, Orsay, F-91405 \\[0.5cm]
\email{stephane.ilic@irap.omp.eu, alain.blanchard@irap.omp.eu, marian.douspis@ias.u-psud.fr}}

\abstract{The abundance of clusters of galaxies is known to be a potential source of cosmological constraints through their mass function. In the present work, we examine the information that can be obtained from the temperature distribution function of X-ray clusters. For this purpose, the mass-temperature ($M$-$T$) relation and its statistical properties are critical ingredients. Using a combination of cosmic microwave background (CMB) data from Planck and our estimations of X-ray cluster abundances, we use Markov chain Monte Carlo (MCMC) techniques to estimate the $\Lambda$CDM cosmological parameters and the mass to X-ray temperature scaling relation simultaneously. We determine the integrated X-ray temperature function of local clusters using flux-limited surveys. A local comprehensive sample was build from the BAX X-ray cluster database, allowing us to estimate the local temperature distribution function above $\sim$1 keV. We model the expected temperature function from the mass function and the $M$-$T$ scaling relation. We then estimate  the cosmological parameters and the parameters of the $M$-$T$ relation (calibration and slope) simultaneously. The measured temperature function of local clusters in the range $\sim\!\!1$-$10$ keV is well reproduced once the calibration of the $M$-$T$ relation is treated as a free parameter, and therefore is  self-consistent with respect to the $\Lambda$CDM cosmology. The best-fit values of the standard cosmological parameters as well as their uncertainties are unchanged by the addition of clusters data. The calibration of the mass temperature relation, as well as its slope, are determined with $\sim10\%$ statistical uncertainties. This calibration leads to masses that are $\sim\!\!75\%$ larger than X-ray masses used in Planck.}

\maketitle

%%%%%%%%%%%%%%%%%%%%%%%%%%%%%%%%%%%%%%%%%%%%%%%%%%%%%%%%%%%%%%%%%%%%%%%%%%%%%%%%%%%%%%%%%%%%%%%%%%%%%%%%%%%%%%%%%%%%%%%%%%%%%%%%%%%%%%%%%%%%%%%%%%%%%%%%%%%%%%%%%%%%%%%%%%
%%%%%%%%%%%%%%%%%%%%%%%%%%%%%%%%%%%%%%%%%%%%%%%%%%%%%%%%%%%%%%%%%%%%%%%%%%%%%%%%%%%%%%%%%%%%%%%%%%%%%%%%%%%%%%%%%%%%%%%%%%%%%%%%%%%%%%%%%%%%%%%%%%%%%%%%%%%%%%%%%%%%%%%%%%
%%%%%%%%%%%%%%%%%%%%%%%%%%%%%%%%%%%%%%%%%%%%%%%%%%%%%%%%%%%%%%%%%%%%%%%%%%%%%%%%%%%%%%%%%%%%%%%%%%%%%%%%%%%%%%%%%%%%%%%%%%%%%%%%%%%%%%%%%%%%%%%%%%%%%%%%%%%%%%%%%%%%%%%%%%

\section{Introduction}
\label{sec:intro}

The quest for cosmological parameters has made dramatic progresses in recent years. The most spectacular progress has come from the evidence for accelerated expansion from the Hubble diagram of distant supernovae \citep{Riess98, Perlmutter99}. Although astrophysical evolution of the type Ia supernovae would easily explain their Hubble diagram \citep{Drell2000,2002astro.ph..1196W,fbz2009}, the measurements of the angular power spectrum of the cosmic microwave background (CMB) fluctuations and of the power spectrum of galaxy distribution on large scales have enabled the confirmation of several predictions of the dark energy dominated, cold dark matter picture \citep{2010A&ARv..18..595B,2013PhR...530...87W}. The parameters of this model have now been estimated with impressive precision \citep{Planck2013_XVI}, making alternative explanations rather contrived. The actual origin of the acceleration now appears as one of the most intriguing and challenging problem of modern cosmology and fundamental physics. While the simplest explanation is an Einstein cosmological constant presently dominating the energy density of the Universe, various options have been investigated in which the dark energy behaves like a fluid with its equation of state $w$, parametrised by $p = w\, \rho$ ($\rho$ being the pressure and $p$ the pressure of the fluid). For these so-called ``quintessence" models, the dark energy is due to the presence of a scalar field, the evolution of its potential determining its equation of state. Modifying the Lagrangian of the field opens an almost unlimited range of possibilities for the properties of the dark energy fluid \citep{2006IJMPD..15.1753C}. However, another explanation for the origin of the cosmic acceleration comes from the modification of the law of gravitation at large scales, i.e. a deviation from general relativity at cosmological scales \citep{2010AnPhy.325.1479J,Eth2013}. In this kind of  case, the dynamics of the expansion could be identical to that of a concordant model and all geometrical tests  provide cosmological parameters identical to those inferred in a Friedman-Lema\^{\i}tre model, but the gravitational dynamics of matter fluctuations may differ, offering a possible signature of the modified gravity. There is, therefore, a specific interest in measuring the gravitational growing rate of fluctuations. The evolution of the abundance of clusters with redshift is exponentially sensitive to the growing rate of linear fluctuations \citep{bb98}. This has motivated the use of cluster abundance evolution as a cosmological test \citep{ob92,1997ApJ...489L...1H,1998MNRAS.298.1145E,1999MNRAS.303..535V,2001ApJ...561...13B,2002ARA&A..40..539R,2005RvMP...77..207V,2014PhyU...57..317V,Planck2013_XX}. X-ray instruments, such as HEAO1, Einstein, ROSAT, XMM, Chandra have enabled the construction of flux-limited surveys of X-ray clusters, the luminosity and temperature of which can be measured accurately. However, conclusions obtained from the analysis of these samples of X-ray clusters have lead to somewhat contradictory results. While some early works based on distant X-ray clusters  \citep{ob97} indicated a significantly lower abundance of clusters at redshift $0.3-0.5$, i.e. a sign of a high-density universe (\citealt{1998A&A...329...21S}, \citealt{1999ApJ...517...40B}, \citealt{1999MNRAS.303..535V}, \citealt{1999ApJ...518..521R}), some recent analyses tend to favour a lower density concordance cosmology \citep{2000ApJ...534..565H,2001ApJ...561...13B,2009ApJ...692.1060V}. On the other hand, \citet{Vauclair2003} showed that the abundance in all existing ROSAT cluster surveys were consistently reproduced by Einstein de Sitter models, which were normalised to the local temperature distribution function using the XMM luminosity temperature evolution of X-ray clusters \citep{Lumb2004}. This result was shown to be insensitive to the calibration of the mass-temperature relation, but relies on standard scaling relations for its evolution. Alternatively, assuming that the mass-temperature relation evolves significantly from the standard scaling law, the X-ray cluster distribution was found to agree well with the concordance cosmology with an appropriate level of evolution. 
More recently, the Sunyaev-Zel'dovitch (SZ) effect has been used in microwave observations to build large samples of galaxy clusters \citep{Hass13,Reichardt2012,Planck2013_XXIX}. Constraints obtained from cosmological samples of few hundred of objects are not limited by low  number statistics, but by the calibration of the scaling laws \citep[][hereafter PXX]{Planck2013_XX}. Depending on the normalisation of the $Y$-$M$ scaling relation, the constraints go from low matter density and low $\sigma_8$  (cf. \citetalias{Planck2013_XX}) to higher values consistent with Planck CMB results \citep{Mantz2014}.

In the present paper, we address the problem of the calibration of the mass-temperature relation of X-ray clusters, using their distribution function at low redshift. We perform a Monte-Carlo Markov Chain (MCMC) analysis to constrain simultaneously the X-ray cluster scaling law and the standard cosmological parameters, with the help of the Planck CMB data. This allows us to derive the mass-temperature relation of clusters in the $\Lambda$CDM model consistently with the Planck CMB data. We compare our result with the calibration derived from pure X-ray analysis, which was used as a prior in the Planck analysis of SZ counts with Planck.

Our paper is structured as follows: In Sect. \ref{sec:mf} we  discuss  the mass function used in theoretical predictions of cluster abundances. In Sect. \ref{sec:tmpfct} we introduce the clusters sample used in our analysis, and then detail how we construct the theoretical and observed temperature function. Finally, in section \ref{sec:params} we present the results of our MCMC analysis and  associated discussion.

%%%%%%%%%%%%%%%%%%%%%%%%%%%%%%%%%%%%%%%%%%%%%%%%%%%%%%%%%%%%%%%%%%%%%%%%%%%%%%%%%%%%%%%%%%%%%%%%%%%%%%%%%%%%%%%%%%%%%%%%%%%%%%%%%%%%%%%%%%%%%%%%%%%%%%%%%%%%%%%%%%%%%%%%%%
%%%%%%%%%%%%%%%%%%%%%%%%%%%%%%%%%%%%%%%%%%%%%%%%%%%%%%%%%%%%%%%%%%%%%%%%%%%%%%%%%%%%%%%%%%%%%%%%%%%%%%%%%%%%%%%%%%%%%%%%%%%%%%%%%%%%%%%%%%%%%%%%%%%%%%%%%%%%%%%%%%%%%%%%%%
%%%%%%%%%%%%%%%%%%%%%%%%%%%%%%%%%%%%%%%%%%%%%%%%%%%%%%%%%%%%%%%%%%%%%%%%%%%%%%%%%%%%%%%%%%%%%%%%%%%%%%%%%%%%%%%%%%%%%%%%%%%%%%%%%%%%%%%%%%%%%%%%%%%%%%%%%%%%%%%%%%%%%%%%%%

\section{Mass function and abundances}
\label{sec:mf}

The halo mass function, i.e. the distribution function of objects relative to their mass, is widely used for cosmological constraints. A theoretical formalism has been proposed by \citet{Press1974}, which allows us to relate the non-linear mass function of cosmic structures to the linear amplitude of the fluctuations, usually specified by the matter power spectrum $P(k)$. We recall the basic assumptions of the derivation of the mass function as follows: 
\begin{itemize}
\item[$\bullet$] {\it i)} the abundance of non-linear haloes with mass greater than some mass $M$ is essentially related to the probability that an elementary volume of the Universe lies in a region of the {\emph linear } field\footnote{Properly smoothed by a window function with a scale $R$, associated with the mass $M$ enclosed by the window function.} that passes the density threshold $\delta_t$  corresponding to the condition for non-linear collapse, i.e.
\begin{equation}
        \nu_t = \frac{\delta_t}{\sigma(M)} \ ,
\end{equation}
with $\sigma^2(M)$ the variance of the field smoothed by the window function. Usually, this function is a top-hat window in real space, although Gaussian windows (and more rarely others) are occasionally used. Any object that is formed has a contrast density above some (non-linear) threshold $\Delta$, which is taken as a definition for the objects counted by the mass function. 
\end{itemize} 
\begin{itemize} 
\item[$\bullet$] {\it ii)} the aforementioned probability can be written as 
\begin{equation}
        \int_{\nu_t}^{+\infty} g(\nu)\d \nu \ ,
\end{equation}
and has been assumed to be independent of the primordial spectrum, but might depend on the Gaussian or non-Gaussian nature of the initial fluctuations. 
\end{itemize} 
This limited set of generic assumptions leads to an universal self-similar form of the mass function, which can be then computed for arbitrary models \citep{Blanchard1992} without requiring the systematic use of numerical simulations. The mass function can then be derived as 
\begin{equation}\label{eq:nm0}
        \frac{dn}{dM} = \frac{\bar{\rho}}{M}g(\nu)\frac{\d\nu}{\d M}= -\frac{\bar{\rho}}{M^2}\frac{\d \ln \sigma}{\d \ln M} \nu g(\nu) \ ,
\end{equation}
and is commonly expressed as {\citep[see e.g.][]{Jenkins2001}:
\begin{equation}\label{eq:nm}
        \frac{dn}{dM} = - \frac{\bar{\rho}}{M} \frac{1}{\sigma} f(\sigma) \frac{d\sigma}{dM}
,\end{equation}
where $f(\sigma)$ ($=\nu g(\nu)$) is a function that is usually fitted on numerical simulations, and $\bar{\rho}$ is the mean matter density of the Universe\footnote{It can be noticed that the scaling  resulting from \ref{eq:nm} is not identical to those predicted by \ref{eq:nm0} because of the a priori dependence of the threshold $\delta_t$ on cosmological models.}. When using a top-hat filter for the smoothing of the initial fluctuations, the mass $M$ corresponds directly to the mass within a sphere of (comoving) radius $R$,
\begin{equation}
        M = \frac{4\pi}{3}R^3\bar{\rho} \ .
\end{equation}
The non-linear objects that the mass function describes are those that present a (non-linear) density contrast $\Delta$ corresponding to the linear threshold  $\delta_t$, i.e.  
\begin{equation}
        M = \frac{4\pi}{3}R^3_{nl}(1+\Delta)\bar{\rho} \ ,
\end{equation}
where we have $R^3 = R^3_{nl}(1+\Delta)$. Note that the non-linear density contrast can also be defined relative to the critical density:  $\Delta_c = \Delta\Omega_m(z)$. A final simplifying assumption is that the threshold $\delta_t$ and the non-linear contrast density $\Delta$ can be deduced from a simple spherical model.

%%%%%%%%%%%%%%%%%%%%%%%%%%%%%%%%%%%%%%%%%%%%%%%%%%%%%%%%%%%%%%%%%%%%%%%%%%%%%%%%%%%%%%%%%%%%%%%%%%%%%%%%%%%%%%%%%%%%%%%%%%%%%%%%%%%%%%%%%%%%%%%%%%%%%%%%%%%%%%%%%%%%%%%%%%
%%%%%%%%%%%%%%%%%%%%%%%%%%%%%%%%%%%%%%%%%%%%%%%%%%%%%%%%%%%%%%%%%%%%%%%%%%%%%%%%%%%%%%%%%%%%%%%%%%%%%%%%%%%%%%%%%%%%%%%%%%%%%%%%%%%%%%%%%%%%%%%%%%%%%%%%%%%%%%%%%%%%%%%%%%

\subsection{Reliable mass functions}
\label{subsec:rmf}

Although the Press and Schechter approach was established in 1974, it is only after large N-body simulations were available that the validity of an universal self-similar mass function gained in strength \citep{Efstathiou1988}. New versions of the mass function based on extensive numerical simulations have since been developed, greatly improving  the accuracy of the description of the halo mass function. Departures from universality have been claimed when the abundance is examined with an accuracy better than 20\% \citep{2006ApJ...646..881W,2011MNRAS.410.1911C}.  We used two types of descriptions of the mass function, and in particular of the fitting function $f$ from Eq.~(\ref{eq:nm}). The first variant of the function $f$ considered is that obtained by \citet[SMT hereafter]{Sheth2001}, while the second is from \citet{Tinker2008}. 
As we  show in what follows, the difference between these mass functions is appreciable but not critical given the statistical uncertainties in current samples of X-ray clusters; nevertheless, the difference could be as large as the (1 $\sigma$) errors in measured abundances. Furthermore, although the \citet{Tinker2008} mass function produces fewer structures than the SMT mass function, the recent and detailed investigations of \citet{Crocce2010} concludes that the SMT mass function seems to {\em{\textup{underestimate}}} the actual abundance of cosmic structures, a claim comforted by \citet{Bhattacharya2011} even if a comprehensive comparison is delicate.  A further source of uncertainty comes from the fact that current estimations of the mass function are derived from numerical simulations in which the gas physics is not included, whereas it may have an appreciable effect \citep[see e.g.][]{Cui2012}. 

We therefore compare the results derived from both the SMT and the Tinker mass functions. Any difference that arises is thus  considered  an estimation of the magnitude of systematics uncertainties that may remain because of the lack of knowledge of the theoretical mass function.
The SMT expression for the mass function, which is evaluated for a top-hat filtering using virial quantities, is given by 
\begin{equation}\label{eq:SMT}
        f_{\rm ST}(\sigma) =A\sqrt{\frac{2a}{\pi}}\left[1+\left(\frac{\sigma^2}{a\delta_c^2}\right)^p\right]\frac{\delta_c}{\sigma}\exp\left[-\frac{a\delta_c^2}{2\sigma^2}\right]
\end{equation}
where $A=0.3222$, $a=0.707$, $p=0.3$, as determined after fitting on N-body simulations results. The parameter $\delta_c \simeq 1.686$ is the critical over-density for spherical collapse. The \citet{Tinker2008} mass function, on the other hand, reads \begin{equation}\label{eq:Tkmassfunction}
        f_{\rm T}(\sigma,z) =A\left(\left(\frac{b}{\sigma}\right)^a + 1\right) \exp\left[-\frac{c}{\sigma^2}\right]
,\end{equation}
where $A$, $a$, $b,$ and $c$ are also fitted parameters, but are functions of the redshift and the over-density contrast $\Delta$ \citep[see][for more details]{Tinker2008}. 

To illustrate the difference between the two mass functions, we  calculated the haloes mass distribution functions for the best-fit cosmology as determined by Planck. Figure \ref{fig:mass_fcts_comp} shows the two mass functions at a local redshift of $z=0.05$. In the presented cluster mass range (above $10^{14}$ M$_\odot$), the difference between the two distributions increases with mass, reaching roughly a factor of two at a mass of $\sim 3\times10^{15}$M$_\odot$. We computed  the Tinker mass function again, increasing the value of $\sigma_8$ from 0.83 to 0.87. The difference between the mass functions then becomes very small, and the relative evolution behaves very similarly. This is an important point to notice, as it implies that constraints on the evolution of the clusters gas physics are almost independent of the choice of the mass function, while the uncertainty in the mass function is expected to make a noticeable difference in the estimation of the amplitude of matter fluctuations, with negligible impact on other parameters (at the level of present-day quality data).

\begin{figure}[t]
\centering
\includegraphics[width=\columnwidth]{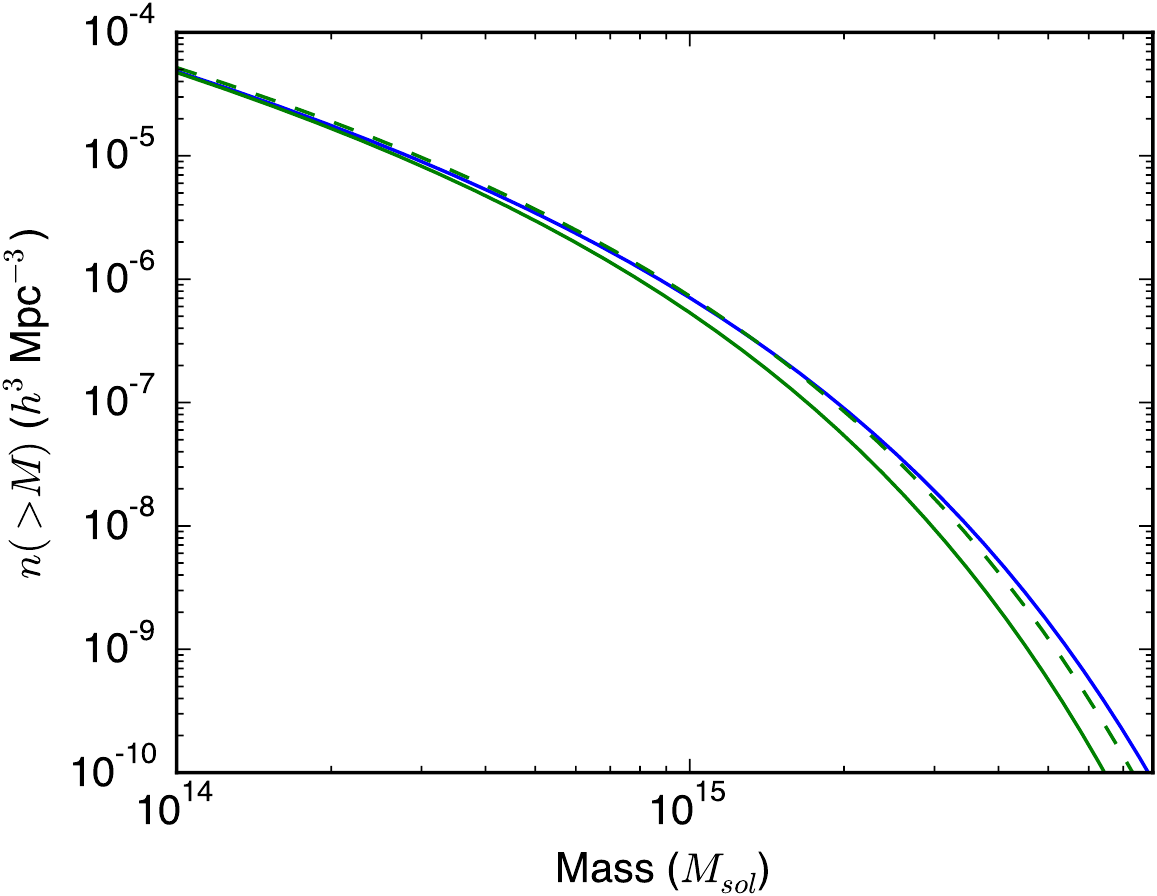}
\caption{Comparison of integrated mass functions computed for the \citet[][blue solid line]{Sheth2001} and the \citet[][green solid line]{Tinker2008} fitting function $f(\sigma)$, in the Planck best-fit cosmology. The green dashed line corresponds to the Tinker mass function with a $\sigma_8=0.87$ instead of the best-fit value of 0.83. All mass functions are estimated at virial mass and redshift of 0.05.}
\label{fig:mass_fcts_comp}
\end{figure}

%%%%%%%%%%%%%%%%%%%%%%%%%%%%%%%%%%%%%%%%%%%%%%%%%%%%%%%%%%%%%%%%%%%%%%%%%%%%%%%%%%%%%%%%%%%%%%%%%%%%%%%%%%%%%%%%%%%%%%%%%%%%%%%%%%%%%%%%%%%%%%%%%%%%%%%%%%%%%%%%%%%%%%%%%%
%%%%%%%%%%%%%%%%%%%%%%%%%%%%%%%%%%%%%%%%%%%%%%%%%%%%%%%%%%%%%%%%%%%%%%%%%%%%%%%%%%%%%%%%%%%%%%%%%%%%%%%%%%%%%%%%%%%%%%%%%%%%%%%%%%%%%%%%%%%%%%%%%%%%%%%%%%%%%%%%%%%%%%%%%%

\subsection{Mass function and recent cosmological constraints}
\label{subsec:mfandcosmo}

The mass function \citep{1992ApJ...386L..33L} is sensitive to the cosmological parameters, especially the matter density parameter $\Omega_m$ and the present-day amplitude of the fluctuations $\sigma_8$. In scenarios such as the $\Lambda$ dominated cold dark matter ($\Lambda$CDM) paradigm, the matter power spectrum  is well defined, and depends on a few parameters that can be constrained from the three standard cosmological data sets: the cosmic microwave background (CMB), the type Ia Supernovae, and the galaxy power spectrum.

In the following, the determination of the cosmological parameters as well as the parameters describing cluster physics are performed by combining the CMB data from Planck \citep[][more precisely, the so-called Planck$+$WP data combination described in the paper]{Planck2013_I} and cluster abundance data. The estimations of the parameters are obtained through a MCMC analysis using the COSMOMC package \citep{Lewis2002,Lewis2013}. Although we could in principle combine CMB data with other sets of data, such as the SNIa Hubble diagram or the BAO signal, this leads to essentially identical cosmological parameters; we therefore choose to keep only Planck CMB data to allow a more direct comparison with Planck clusters results. We then estimate the range of predicted abundances for models in which parameters fall within the 68\% range. The results and the 68\% error bars  on cosmological parameters are given in the first column of Table \ref{tab:param_MCMC}.

Figure \ref{fig:mass_fcts_contours} shows the $1\sigma$ uncertainty region for the expected mass function determined for $\Delta_c = 500$. We compared this result with the determination from \citep{Vikhlinin2009a}: their reference model is shown with the dashed line; their cosmological parameters are $\Omega_m=0.3$, $h=0.72$ and $\sigma_8=0.746$. It is clear that the expected mass distribution function in $\Lambda$CDM with parameters and uncertainties determined from standard cosmological constraints leads to a much larger overall abundance of clusters. This is clearly related to the temperature mass relation used in \citep{Vikhlinin2009a}. We follow a different approach: we assume a standard $\Lambda$CDM model and fit  the CMB data and the abundance of X-ray clusters simultaneously to infer constraints on the mass temperature relation required for consistency. This  allows us to make a comparison with masses estimated by more observational approaches and examine what are the consequences for the interpretation of Planck SZ clusters counts.

\begin{figure}[ht!]
\includegraphics[width=\columnwidth]{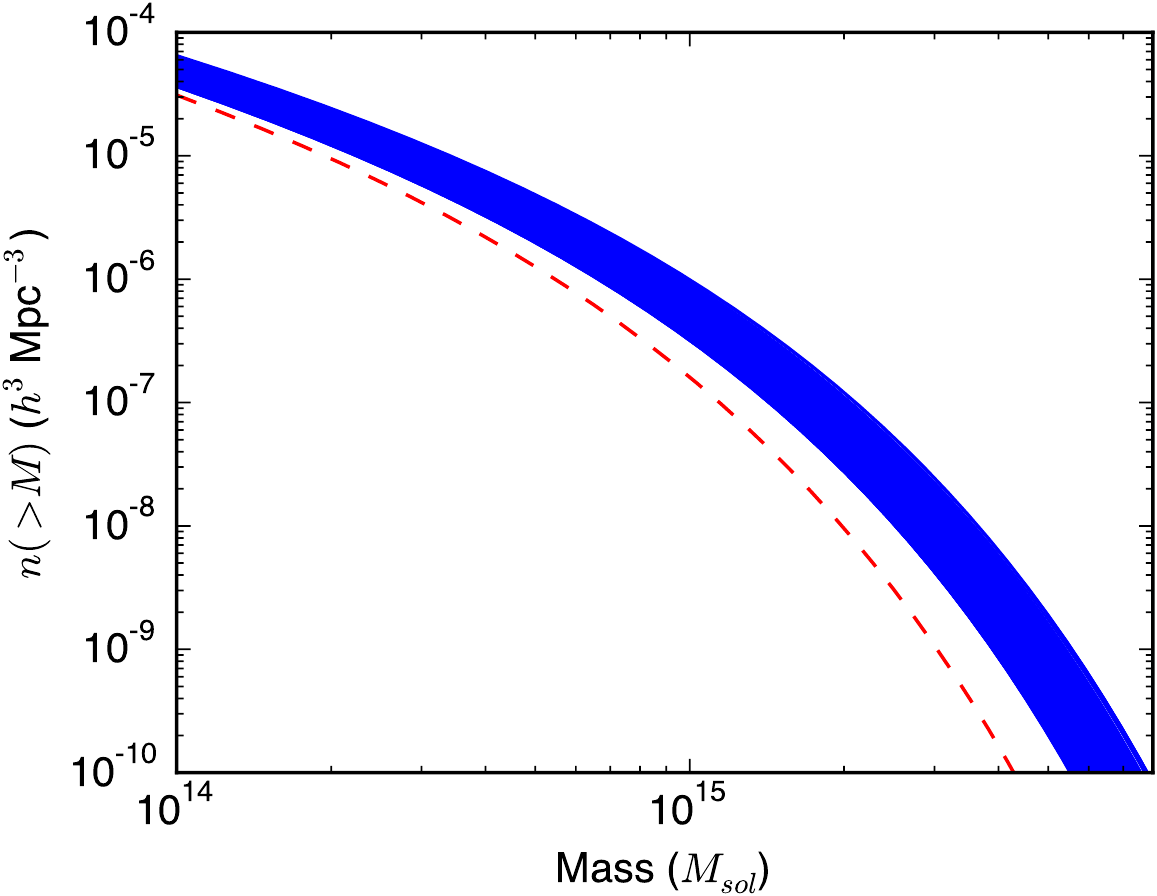}
\caption{In  blue, $1\,\sigma$ region for the Tinker mass function according to current cosmological constraints (CMB measurements). The red dashed line represents the mass function of \cite{Vikhlinin2009a} with $\Omega_m=0.3$, $h=0.72$ and $\sigma_8=0.746$. All mass functions are computed at $z=0.05$.}
\label{fig:mass_fcts_contours}
\end{figure}

%%%%%%%%%%%%%%%%%%%%%%%%%%%%%%%%%%%%%%%%%%%%%%%%%%%%%%%%%%%%%%%%%%%%%%%%%%%%%%%%%%%%%%%%%%%%%%%%%%%%%%%%%%%%%%%%%%%%%%%%%%%%%%%%%%%%%%%%%%%%%%%%%%%%%%%%%%%%%%%%%%%%%%%%%%
%%%%%%%%%%%%%%%%%%%%%%%%%%%%%%%%%%%%%%%%%%%%%%%%%%%%%%%%%%%%%%%%%%%%%%%%%%%%%%%%%%%%%%%%%%%%%%%%%%%%%%%%%%%%%%%%%%%%%%%%%%%%%%%%%%%%%%%%%%%%%%%%%%%%%%%%%%%%%%%%%%%%%%%%%%
%%%%%%%%%%%%%%%%%%%%%%%%%%%%%%%%%%%%%%%%%%%%%%%%%%%%%%%%%%%%%%%%%%%%%%%%%%%%%%%%%%%%%%%%%%%%%%%%%%%%%%%%%%%%%%%%%%%%%%%%%%%%%%%%%%%%%%%%%%%%%%%%%%%%%%%%%%%%%%%%%%%%%%%%%%

\section{Determination of the temperature distribution function}
\label{sec:tmpfct}

Existing surveys of X-ray clusters provide comprehensive sets of clusters for which the selection function is reasonably well understood. When the temperature is known for every cluster of a given survey, we can estimate the temperature distribution function at the typical redshift of the survey. The match between the mass function and temperature distribution function provide self-consistent information about clusters for $\Lambda$CDM models.

%%%%%%%%%%%%%%%%%%%%%%%%%%%%%%%%%%%%%%%%%%%%%%%%%%%%%%%%%%%%%%%%%%%%%%%%%%%%%%%%%%%%%%%%%%%%%%%%%%%%%%%%%%%%%%%%%%%%%%%%%%%%%%%%%%%%%%%%%%%%%%%%%%%%%%%%%%%%%%%%%%%%%%%%%%
%%%%%%%%%%%%%%%%%%%%%%%%%%%%%%%%%%%%%%%%%%%%%%%%%%%%%%%%%%%%%%%%%%%%%%%%%%%%%%%%%%%%%%%%%%%%%%%%%%%%%%%%%%%%%%%%%%%%%%%%%%%%%%%%%%%%%%%%%%%%%%%%%%%%%%%%%%%%%%%%%%%%%%%%%%

\subsection{A new complete sample of low-redshift clusters}
\label{subsec:lowzclus}

We built a new sample of X-ray selected clusters, with a strategy that is essentially the same as in \cite{Blanchard2000}, who published the first temperature distribution function based on a sample selected from ROSAT fluxes. Currently, many measurements of clusters fluxes are often available and of good quality, coming from different instruments and sometimes estimated in different bands. It is preferable to have samples with flux estimations that are as homogeneous as possible. However, this is a rather complicated task: for instance, the conversion between counts and flux requires a spectral model of the source. This is generally done assuming a single temperature, which may not be available at the time of the flux measurement. Furthermore, the emissivity of X-ray clusters cannot be represented by a single temperature: occasionally, the central part of clusters are complex multi-components temperature systems, such as occurs in cooling flow clusters. For this reason, the emissivity in the central region is often corrected. We built our sample from the online database BAX\footnote{``Base de Donn\'ees Amas de Galaxies X'', \url{http://bax.irap.omp.eu}} \citep{Sadat2004}, which provides published fluxes that are homogenised to the same energy band. Therefore, for our selection we used the BAX working energy band corresponding to the ROSAT [0.1-2.4] keV energy band. We removed clusters that are too close to the galactic plane by discarding objects between $20 < b < -20$, where $b$ is the latitude in the galactic coordinate system; this results in a sky coverage of $7.96\rm\ sr$ ($\sim63\%$ of the sky). We did not apply any further geometrical selection. To limit the internal evolution of the abundance of objects, we restricted the redshift range of our sample and did not include clusters with redshift above $z=0.1$. 

To compute the detection volume for all selected clusters, we need a sample with a well-understood selection function. In our selection criterion, we limited our sample to objects known as \emph{\textup{bona fide}} clusters brighter than a minimal flux, such that at least one temperature measurement has been published for every selected cluster. In practice, this sets the minimal flux to $f_{min}=1.810^{-11} \rm{erg.s^{-1}.cm^{-2}}$, at which the whole selected area is complete from the various ROSAT samples. We note that BAX contains objects brighter than this flux, which are no longer considered  clusters  and were removed from the final sample. With our selection criteria combined, we finally obtained a sample of 73 clusters covering a temperature range of $[0.8-9]$ keV with a mean redshift of $z\sim 0.05$, making this sample the largest ever  used for the determination of the local temperature distribution function (with actual temperature measurements).

\begin{table*}[ht]
%\begin{center}
   \begin{tabular}{lllll}
      \hline
     \hline
Cluster & z &$T_{X}$$^{a}$ &  $s_{X}$$^{b}$ &  Ref$^{c}$.\\
      \hline
 NGC4552 & 0.001 & 0.8$_{ -0.05}^{ +0.05 }$ &  19.1 & {\sl 1,1 } \\
NGC4343GROUP & 0.004 & 0.9$_{ -0.02}^{ +0.02 }$ &  39.7 & {\sl 5,2 } \\
VIRGOCLUSTER & 0.004 & 2.28$_{ -0.02}^{ +0.02 }$ &  761.6 & {\sl 3,3 } \\
VIRGOS & 0.004 & 1.13$_{ -0.02}^{ +0.02 }$ &  18.51 & {\sl 1,4 } \\
NGC1399GROUP & 0.005 & 1.56$_{ -0.04}^{ +0.03 }$ &  86.8 & {\sl 2,2 } \\
NGC5044GROUP & 0.008 & 1.22$_{ -0.04}^{ +0.03 }$ &  58.1 & {\sl 5,6 } \\
ABELL3526 & 0.011 & 3.92$_{ -0.02}^{ +0.02 }$ &  271.89 & {\sl 1,6 } \\
NGC1550GROUP & 0.012 & 1.34 $_{ -0.00}^{ +0.00 }$ &  31.2 & {\sl 5,8 } \\
RXCJ0419.6+0224 & 0.012 & 1.29$_{ -0.02}^{ +0.04 }$ &  40.1 & {\sl 9,6 } \\
ABELL1060 & 0.013 & 3.15$_{ -0.03}^{ +0.03 }$ &  99.5 & {\sl 2, 16} \\
ABELL0262 & 0.016 & 2.25$_{ -0.04}^{ +0.04 }$ &  93.5 & {\sl 2,4 } \\
NGC0507GROUP & 0.019 & 1.44$_{ -0.1}^{ +0.08 }$ &  20 & {\sl 5,2 } \\ 
MKW04 & 0.02 & 1.78$_{ -0.09}^{ +0.07 }$ &  22.7 & {\sl 11,4 } \\
ABELL1367 & 0.022 & 3.55$_{ -0.05}^{ +0.05 }$ &  60.5 & {\sl 2,4 } \\
ABELL1656 & 0.023 & 8.25$_{ -0.1}^{ +0.1 }$ &  344.4 & {\sl 12,4 } \\
ABELL3581 & 0.023 & 1.83$_{ -0.01}^{ +0.02 }$ &  33.4 & {\sl 2,4 } \\
ABELL0400 & 0.024 & 2.43$_{ -0.07}^{ +0.08 }$ &  27.8 & {\sl 2,4 } \\
MKW08 & 0.027 & 3.29$_{ -0.13}^{ +0.14 }$ &  25.1 & {\sl 2,2 } \\
RXJ0341.3+1524 & 0.029 & 2.17$_{ -0.06}^{ +0.06 }$ &  20.1 & {\sl 13,2 } \\
ABELL2199 & 0.03 & 4.37$_{ -0.07}^{ +0.07 }$ &  106.4 & {\sl 10,4 } \\
RXCJ2347.7-2808 & 0.03 & 2.61$_{ -0.05}^{ +0.05 }$ &  56.9 & {\sl 14,4 } \\
ABELL2634 & 0.031 & 3.45$_{ -0.1}^{ +0.1 }$ &  24.1 & {\sl 2,4 } \\
ABELL0496 & 0.033 & 4.12$_{ -0.07}^{ +0.07 }$ &  83.3 & {\sl 14,4 } \\
UGC03957CLUSTER & 0.034 & 2.6$_{ -0.04}^{ +0.04 }$ &  19.7 & {\sl 14,4 } \\
ABELL2052 & 0.035 & 3.12$_{ -0.05}^{ +0.06 }$ &  47.1 & {\sl 2,4 } \\
ABELL2063 & 0.035 & 3.57$_{ -0.19}^{ +0.19 }$ &  42.3 & {\sl 14,4 } \\
2A0335+096 & 0.035 & 3.64$_{ -0.05}^{ +0.05 }$ &  91.6 & {\sl 2,4 } \\
ABELL2147 & 0.035 & 4.34$_{ -0.08}^{ +0.07 }$ &  55.2 & {\sl 2,4 } \\
ABELL3571 & 0.039 & 6.81$_{ -0.1}^{ +0.1 }$ &  120.9 & {\sl 14,4 } \\
ABELL0576 & 0.039 & 3.68$_{ -0.11}^{ +0.11 }$ &  30.1 & {\sl 14,4 } \\
EXO0422-086 & 0.04 & 2.84$_{ -0.09}^{ +0.09 }$ &  30.8 & {\sl 14,4 } \\
ABELL2657 & 0.04 & 3.62$_{ -0.15}^{ +0.15 }$ &  25.3 & {\sl 14,4 } \\
ABELL2107 & 0.041 & 4$_{ -0.1}^{ +0.1 }$ &  19.4 & {\sl 15,16 } \\ 
ABELL2589 & 0.041 & 3.38$_{ -0.08}^{ +0.08 }$ &  25.9 & {\sl 2,4 } \\
ABELL0119 & 0.044 & 5.4$_{ -0.23}^{ +0.23 }$ &  40.5 & {\sl 17,4 } \\
MKW03S & 0.045 & 3.45$_{ -0.06}^{ +0.08 }$ &  33.0 & {\sl 11,4 } \\
ABELL3376 & 0.046 & 4.48$_{ -0.12}^{ +0.11 }$ &  24.5 & {\sl 19,4 } \\
\hline
 \hline
\end{tabular}
\hspace*{2mm}
\vspace{0mm}
   \begin{tabular}{lllll}
      \hline
     \hline
Cluster & z &$T_{X}$$^{a}$ &  $s_{X}$$^{b}$ &  Ref$^{c}$.\\
      \hline
ABELL1736 & 0.046 & 2.95$_{ -0.09}^{ +0.09 }$ &  35.4 & {\sl 14,4 } \\
ABELL1644 & 0.047 & 4.86$_{ -0.2}^{ +0.2 }$ &  40.3 & {\sl 17,4 } \\
ABELL3558 & 0.048 & 4.78$_{ -0.13}^{ +0.13 }$ &  67.2 & {\sl 17,4 } \\
ABELL4059 & 0.048 & 4.25$_{ -0.08}^{ +0.08 }$ &  31.7 & {\sl 14,4 } \\
ABELL3562 & 0.049 & 4.31$_{ -0.12}^{ +0.12 }$ &  29.3 & {\sl 14,4 } \\
IC1365 & 0.049 & 3.93$_{ -0.1}^{ +0.1 }$ &  19.0 & {\sl 7,2 } \\
ABELL3391 & 0.051 & 5.39$_{ -0.19}^{ +0.19 }$ &  22.2 & {\sl 14,4 } \\
ABELL3395SW & 0.051 & 4.42$_{ -0.22}^{ +0.22 }$ &  20.1 & {\sl 11,2 } \\
ABELL3395 & 0.051 & 5.1$_{ -0.17}^{ +0.17 }$ &  31.4 & {\sl 14,14 } \\
ABELL0754 & 0.054 & 8.93$_{ -0.24}^{ +0.24 }$ &  33.7 & {\sl 17,4 } \\
ABELL0780 & 0.054 & 3.42$_{ -0.01}^{ +0.01 }$ &  47.8 & {\sl 10,4 } \\
ABELL0085 & 0.055 & 5.78$_{ -0.22}^{ +0.22 }$ &  74.3 & {\sl 17,4 } \\
ABELL3667 & 0.056 & 6.33$_{ -0.06}^{ +0.06 }$ &  72.0 & {\sl 14,4 } \\
RXCJ0102.7-2152 & 0.057 & 4.01$_{ -0.11}^{ +0.11 }$ &  21.2 & {\sl 14,4 } \\
ABELLS1101 & 0.058 & 2.59$_{ -0.01}^{ +0.01 }$ &  24.8 & {\sl 19,4 } \\
ABELL2256 & 0.058 & 6.4$_{ -0.25}^{ +0.25 }$ &  52.2 & {\sl 17,20 } \\
ABELL3266 & 0.059 & 7.46$_{ -0.22}^{ +0.22 }$ &  58.1 & {\sl 17,4 } \\
ABELL3158 & 0.06 & 5$_{ -0.18}^{ +0.18 }$ &  37.9 & {\sl 17,4 } \\
ABELL1795 & 0.062 & 6.12$_{ -0.05}^{ +0.05 }$ &  62.7 & {\sl 21,4 } \\
ABELL0399 & 0.072 & 7.95$_{ -0.31}^{ +0.35 }$ &  32.5 & {\sl 18,4 } \\
ABELL2065 & 0.073 & 5.36$_{ -0.2}^{ +0.2 }$ &  25.0 & {\sl 17,4 } \\
ABELL0401 & 0.074 & 6.37$_{ -0.19}^{ +0.19 }$ &  52.8 & {\sl 18,4 } \\
ZWCL1215.1+0400 & 0.075 & 6.45$_{ -0.27}^{ +0.27 }$ &  21.8 & {\sl 17,4 } \\
ABELL3112 & 0.075 & 5.02$_{ -0.15}^{ +0.15 }$ &  31.0 & {\sl 17,4 } \\
ABELL3822 & 0.076 & 5.12$_{ -0.19}^{ +0.26 }$ &  19.3 & {\sl 2,4 } \\
ZWCL1742.1+3306 & 0.076 & 3.59$_{ -0.04}^{ +0.02 }$ &  18.7 & {\sl 7,2 } \\
ABELL2029 & 0.077 & 7.7$_{ -0.41}^{ +0.41 }$ &  69.4 & {\sl 17,4 } \\
ABELL2061 & 0.078 & 4.3$_{ -0.1}^{ +0.1 }$ &  18.6 & {\sl 7,22 } \\
ABELL2255 & 0.081 & 5.79$_{ -0.15}^{ +0.15 }$ &  20.2 & {\sl 17,4 } \\
ABELL1650 & 0.084 & 5.11$_{ -0.06}^{ +0.06 }$ &  24.0 & {\sl 17,4 } \\
ABELL2597 & 0.085 & 4.05$_{ -0.07}^{ +0.07 }$ &  22.1 & {\sl 5,4 } \\
ABELL1651 & 0.085 & 6.26$_{ -0.27}^{ +0.3 }$ &  25.4 & {\sl 18,4 } \\
ABELL0478 & 0.088 & 7.3$_{ -0.24}^{ +0.26 }$ &  51.5 & {\sl 18,4 } \\
ABELL2142 & 0.091 & 8.65$_{ -0.22}^{ +0.22 }$ &  62.4 & {\sl 23,4 } \\
ABELL2244 & 0.097 & 5.37$_{ -0.12}^{ +0.12 }$ &  21.2 & {\sl 14,4 } \\
ABELL3827 & 0.098 & 6.19$_{ -0.1}^{ +0.1 }$ &  19.8 & {\sl 17,2 } \\
          &       &                     &       &  \\
\hline
       \hline
          \end{tabular}

\noindent $^{a}$ temperature in {\sl keV} and 68\% uncertainties.\\
$^{b}$  flux provided by BAX in 10$^{-12}$erg/s/cm$^2$ in the $0.1-2.4$ {\sl keV} ROSAT band. \\
$^{c}$ the first reference is for temperature, the second for the luminosity.
\noindent \caption{ \label{Table 1}  X-ray temperatures (68\% uncertainties), ROSAT 0.1--2.4 keV fluxes, and redshifts of the 73 X-ray clusters used in the present work.}
${\bf Ref.}$
{\sl (1)}  \citealt{2004PASJ...56..965F}
{\sl (2)}  \citealt{2002A&A...383..773I}
{\sl (3)}  \citealt{2000PASJ...52..153M}
{\sl (4)}  \citealt{2002ApJ...567..716R}
{\sl (5)}  \citealt{2010A&A...513A..37H}
{\sl (6)}  \citealt{2004A&A...425..367B}
{\sl (7)}  \citealt{2009ApJ...690..879S}
{\sl (8)}  \citealt{2003PASJ...55..573K}
{\sl (9)}  \citealt{2004AdSpR..34.2525Y}
{\sl (10)} \citealt{2009A&A...493..409S}
{\sl (11)} \citealt{2007MNRAS.380.1554R}
{\sl (12)} \citealt{2001A&A...365L..67A}
{\sl (13)} \citealt{2009ApJ...693.1142S}
{\sl (14)} \citealt{2009ApJ...692.1033V}
{\sl (15)} \citealt{2006PASJ...58..131F}
{\sl (16)} \citealt{1999ApJ...511...65J}
{\sl (17)} \citealt{PlanckER2011_XI}
{\sl (18)} \citealt{2008ApJ...682..821C}
{\sl (19)} \citealt{2007A&A...465..345D}
{\sl (20)} \citealt{2005ApJ...624L..69F}
{\sl (21)} \citealt{2006ApJ...640..691V}
{\sl (22)} \citealt{2004MNRAS.353.1219M}
{\sl (23)} \citealt{2002ApJ...567..163D}
\end{table*} 

%%%%%%%%%%%%%%%%%%%%%%%%%%%%%%%%%%%%%%%%%%%%%%%%%%%%%%%%%%%%%%%%%%%%%%%%%%%%%%%%%%%%%%%%%%%%%%%%%%%%%%%%%%%%%%%%%%%%%%%%%%%%%%%%%%%%%%%%%%%%%%%%%%%%%%%%%%%%%%%%%%%%%%%%%%
%%%%%%%%%%%%%%%%%%%%%%%%%%%%%%%%%%%%%%%%%%%%%%%%%%%%%%%%%%%%%%%%%%%%%%%%%%%%%%%%%%%%%%%%%%%%%%%%%%%%%%%%%%%%%%%%%%%%%%%%%%%%%%%%%%%%%%%%%%%%%%%%%%%%%%%%%%%%%%%%%%%%%%%%%%

\subsection{Modelling the temperature function}
\label{subsec:modeltmpfct}

An ideal situation would be to have a direct estimation of the mass function to compare with theoretical expectations, however, a direct detection of DM haloes based on their mass is not an easy task. Blind lensing surveys may allow this possibility in the future, such as with the \emph{Euclid} project \citep{Laureijs2011}, but this seems out of reach with our currently available means. We are therefore restricted to cluster surveys based on techniques, such as velocity dispersion, X-ray, or SZ signal, and we then apply a relation between the observable and the mass of each cluster present in the survey. It is therefore necessary to have a good understanding of this mass-observable relation, in addition to a selection function that is well under control.

We focus on X-ray clusters and start with the assumption that clusters are self-similar at some level, i.e. that their observable quantities (e.g. luminosity or temperature) can be related to their mass through a simple expression, generally a power law. Hereafter, we call these relations ``scaling relations". \citet{Kaiser1991} proposed that the dependence of these relations on mass and redshift can be inferred from a simple collapse model; we  refer to these as standard scaling relations. Beyond these, it is important to take  the effect of dispersion into account. In the following, we work with  the virial mass (unless specified otherwise), i.e. the mass inside the radius that corresponds to a contrast density equal to $\Delta_v$ in the standard spherical model. Hence, the virial radius  is  
\begin{equation}
        R_v=\frac{1}{1+z}\left(\frac{M}{4\pi /3(1+\Delta_v)\Omega_m\rho_c}\right)^{1/3}
.\end{equation}
To derive the gas temperature, provided the gravitational collapse is the unique source of gas heating, we can assume that the average thermal energy of a particle is related to its average kinetic energy in the gravitational potential of the cluster (incomplete virialisation is accommodated by this assumption), 
\begin{equation}
        kT=f_T \frac\mu m_p\sigma^2
,\end{equation}
with $\mu$ the average molecular weight of the intracluster plasma, $\sigma$ the (1D) velocity dispersion along the line of sight $\sigma~=~1~/\sqrt{3}\sigma_{3D}$, and $f_T$ a parameter accommodating for possible incomplete virialisation. For a single isothermal sphere, the velocity dispersion is 
\begin{equation}
        \sigma^2=GM/R
,\end{equation}
allowing us to derive the (theoretical) scaling law between temperature and mass, i.e.
\begin{equation}\label{eq:MTrelation}
        T=A_{TM}(hM_v)^{2/3}\left(\frac{\Omega_{m} \Delta(\Omega_m,z)}{178}\right)^{1/3}(1+z)^{1+\alpha_{TM}}
,\end{equation}
with $A_{TM}$ the normalisation constant, $M_v$ the virial mass of the cluster expressed in unit of $10^{15}$ solar mass, and $\Delta(\Omega_{m},z)$ the density contrast with respect to the total matter density of the Universe. The $\alpha_{TM}$ parameter is introduced to account for a possible evolution of the scaling law with redshift;  we do not consider this additional parameter and fix $\alpha_{TM}=0$, since our local sample of clusters (all with redshifts lower than 0.1) is not able to constrain this parameter.

Some of the previous works in the literature used  the $L-M$ relation to derive cosmological constraints directly. The main advantage of this method is that the luminosity is much easier to obtain than temperature. As a consequence larger samples of clusters can be used with this kind of  approach. However, the luminosity is proportional to the square of the gas density, which is concentrated in the core of the cluster and therefore less likely to follow a scaling relation with the total mass. Additionally, the $L-T$ relation is known {\em \textup{not to follow}} a standard scaling law, which is generally attributed to the sensitivity of the gas distribution to non-gravitational gas physics. If the $L-T$ relation is not standard then it also implies that luminosity does not follow a standard scaling with mass.

For a given $T-M$ scaling relation assumed with no dispersion, the number of objects that are hotter than a temperature threshold is equivalent to the number of objects that are heavier than a corresponding mass threshold. The theoretical temperature function can be written as
\begin{equation}\label{eq:temp_fct1}
    n(>T) = \int_{T}^{+\infty}\frac{dn}{dt}(t)dt =  \int_{M(T)}^{+\infty}\frac{dn}{dm}(m)dm % why the _sh ??
,\end{equation}
where $M(T)$ is given by the scaling relation (\ref{eq:MTrelation}). The last equation allows us to compute the integrated temperature function from the mass function, i.e. the numerical density of clusters that are hotter than a given temperature.

We have to take the effect of the dispersion in the $T-M$ scaling relation  into account. In the absence of any dispersion, the equation (\ref{eq:temp_fct1}) can be rewritten as follows:
\begin{equation}\label{eq:temp_fct2}
        n(>T) = \int_{0}^{+\infty}\theta(m-M(T))\frac{dn}{dm}(m)dm
,\end{equation}
where $\theta(x)$ is the Heaviside function. In practice, the threshold should be replaced by a smooth function raising from zero to one. Assuming a log-normal dispersion in temperature with a variance $\sigma_T^2$, the observed temperature distribution function can be rewritten as~
\begin{equation}\label{eq:temp_fct3}
        n(>T) = \int_{0}^{+\infty}\frac{1}{2}\left(1.+{\rm erf}(-\frac{(\ln(t)  - \ln (T))^2}{2\sigma_T^2})\right)\frac{dn}{dt}(t)dt
.\end{equation}
For a power-law distribution $n(>T) = K T^{-\alpha} (\alpha > 0)$, the dispersion causes an increase of the observed number of clusters 
\begin{equation}
   \tilde{n}(>T) = K \left(T\left(1-\left(\alpha-1\right)\sigma_T^2/2\right)\right)^{-\alpha} \ ,
\end{equation}
i.e. the effect of dispersion is totally absorbed by a shift in the value of the normalisation of the mass temperature relation 
\begin{equation}
        A_{TM}^{obs} = A_{TM}^{true}/\left(1-\left(\alpha-1\right)\sigma_T^2/2 \right)\ .
\end{equation}
The effective value of $\alpha$ in the range $3-10$keV is of the order of 3 (cf. Fig.~\ref{fig:n_T_fit_data}), so that a 20\% dispersion leads to a measured value of $A_{TM}$ 4\% higher than the ``true'' underlying value of $A_{TM}$. Numerical simulations are indicative of a dispersion of this order. However, as the dispersion is not directly accessible with observations, we do not attempt to correct the inferred value of $A_{TM}$ in the following. The value we derive is therefore be effective, and possibly biased by dispersion.

%%%%%%%%%%%%%%%%%%%%%%%%%%%%%%%%%%%%%%%%%%%%%%%%%%%%%%%%%%%%%%%%%%%%%%%%%%%%%%%%%%%%%%%%%%%%%%%%%%%%%%%%%%%%%%%%%%%%%%%%%%%%%%%%%%%%%%%%%%%%%%%%%%%%%%%%%%%%%%%%%%%%%%%%%%
%%%%%%%%%%%%%%%%%%%%%%%%%%%%%%%%%%%%%%%%%%%%%%%%%%%%%%%%%%%%%%%%%%%%%%%%%%%%%%%%%%%%%%%%%%%%%%%%%%%%%%%%%%%%%%%%%%%%%%%%%%%%%%%%%%%%%%%%%%%%%%%%%%%%%%%%%%%%%%%%%%%%%%%%%%

\subsection{The observed temperature function}
\label{subsec:tmpfctobs}

The value of $A_{TM}$ is a quantity providing a first constraint on clusters gas physics. It can be obtained by comparing an observed temperature distribution function with the expectation from some fiducial model. In this section, we describe how we estimate this temperature distribution function from existing samples of X-ray clusters as well as the uncertainty in these estimations.

The number of clusters $N$ inside a luminosity range $[L,L+\Delta L]$ is the realisation of a stochastic Poisson process with a mean value $\Phi(L)V(L)\Delta(L)$, with $V(L)$ being the volume of detection for an object with luminosity $L$ and $\Phi(L)$ is the luminosity distribution function. For a selection function specified by a flux dependent area $A(f_x,z)$, the search (comoving) volume corresponding to a cluster of luminosity $L$ can be written as 
\begin{equation}\label{eq:VLHenry1}
        V(L)=\int_{0}^{z_{max}}dzA(L,z)\frac{d^2V}{d\Omega dz}
,\end{equation}
where $\frac{dV^2}{d\Omega dz} $ is the comoving volume element per unit area on the sky. By definition, $N/V(L)$ is then an unbiased estimation of $\Phi(L)\Delta L$ and the integrated luminosity function can be written 
\begin{equation}\label{eq:NL}
        \Phi(>L)=\sum_{L_i>L}\frac{1}{V(L_i)}
,\end{equation}
where the sum is performed over all clusters in the sample satisfying the criteria. In our case, we are interested in the temperature distribution function. We can therefore define in a similar fashion a volume search for an object with temperature $T$   , i.e. 
\begin{equation}
        V(T)=\int_{z_{min}}^{z_{max}}A(T,z)\frac{d^2V}{d\Omega dz} \ . 
\end{equation}
The temperature function is given by the following unbiased estimator:
\begin{equation}\label{eq:NT}
        n(>T)=\sum_{T_i>T}\frac{1}{V(T_i)}
\end{equation}
However, in practice the selection function $A(kT,z)$ is not known. Attempts to estimate the volume search $V(kT)$ from the temperature-luminosity relation (and from the search volumes $V(L)$) suffer from a systematic uncertainty due to the fact that the properties of the luminosity-temperature relation is not perfectly known.

For a given flux limit, the maximum redshift of detection $z_{max}$ of a cluster can be determined from its actual flux. To accomplish  this, the flux is first converted into a rest-frame luminosity in the same band of observation, [0.1-2.4] keV in the present case. The maximum redshift of detection is therefore the redshift at which the flux of the cluster reaches the flux threshold. Note that this requires us to evaluate the rest-frame luminosity in the band corresponding to the observation. In practice, we use a Raymond-Smith code for this. The flux is then~
\begin{equation}\label{eq:lx2fx}
        f_x=\frac{L_x}{4\pi d_L(z)^2}
,\end{equation}
with $L_x$ the luminosity in the appropriate rest-frame band, i.e. [0.1(1+z),2.4(1+z)] keV, and $d_L(z)$ the cosmological luminosity distance at a redshift z. 

Figure (\ref{fig:n_T_fit_data}) shows the observed temperature function for local clusters, defined as those with a redshift below 0.1.

\begin{figure}[ht!]
\centering
\includegraphics[width=\columnwidth]{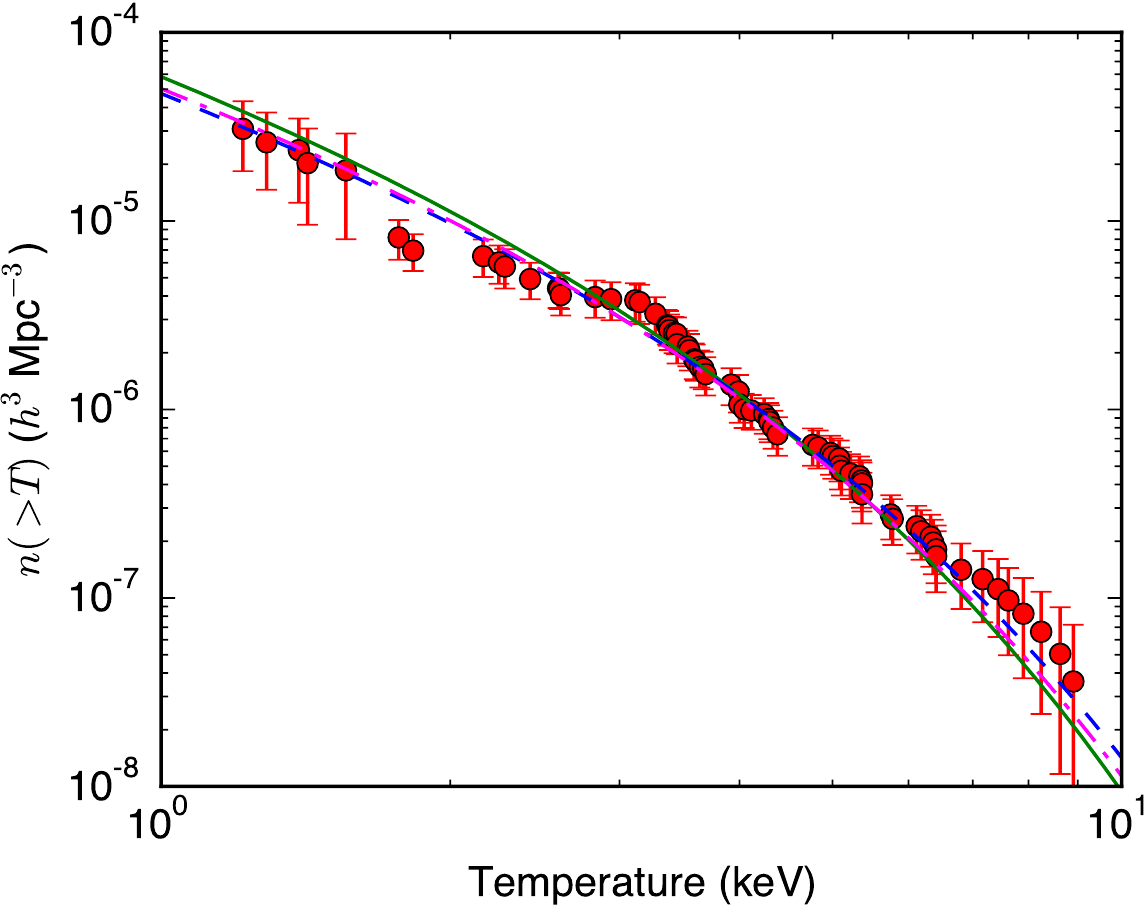}
\caption{Observed temperature function of local clusters (red dots), with error bars defined as the sum of the inverse-square detection volumes of the clusters. The best-fit theoretical models are also shown, in the case of the mass function of SMT (blue dashed curve), and the mass function of \citet{Tinker2008}, with and without a fixed index $\beta_{TM}$ (defined in Eq.~\ref{eq:newMTrelation}) for the $M$-$T$ scaling law ( solid green and dot-dashed purple curves, respectively).}
\label{fig:n_T_fit_data}
\end{figure}

%%%%%%%%%%%%%%%%%%%%%%%%%%%%%%%%%%%%%%%%%%%%%%%%%%%%%%%%%%%%%%%%%%%%%%%%%%%%%%%%%%%%%%%%%%%%%%%%%%%%%%%%%%%%%%%%%%%%%%%%%%%%%%%%%%%%%%%%%%%%%%%%%%%%%%%%%%%%%%%%%%%%%%%%%%
%%%%%%%%%%%%%%%%%%%%%%%%%%%%%%%%%%%%%%%%%%%%%%%%%%%%%%%%%%%%%%%%%%%%%%%%%%%%%%%%%%%%%%%%%%%%%%%%%%%%%%%%%%%%%%%%%%%%%%%%%%%%%%%%%%%%%%%%%%%%%%%%%%%%%%%%%%%%%%%%%%%%%%%%%%
%%%%%%%%%%%%%%%%%%%%%%%%%%%%%%%%%%%%%%%%%%%%%%%%%%%%%%%%%%%%%%%%%%%%%%%%%%%%%%%%%%%%%%%%%%%%%%%%%%%%%%%%%%%%%%%%%%%%%%%%%%%%%%%%%%%%%%%%%%%%%%%%%%%%%%%%%%%%%%%%%%%%%%%%%%

\section{Parameters determination}
\label{sec:params}
  
%%%%%%%%%%%%%%%%%%%%%%%%%%%%%%%%%%%%%%%%%%%%%%%%%%%%%%%%%%%%%%%%%%%%%%%%%%%%%%%%%%%%%%%%%%%%%%%%%%%%%%%%%%%%%%%%%%%%%%%%%%%%%%%%%%%%%%%%%%%%%%%%%%%%%%%%%%%%%%%%%%%%%%%%%%
%%%%%%%%%%%%%%%%%%%%%%%%%%%%%%%%%%%%%%%%%%%%%%%%%%%%%%%%%%%%%%%%%%%%%%%%%%%%%%%%%%%%%%%%%%%%%%%%%%%%%%%%%%%%%%%%%%%%%%%%%%%%%%%%%%%%%%%%%%%%%%%%%%%%%%%%%%%%%%%%%%%%%%%%%%

\subsection{The likelihood analysis}
\label{subsec:like}

The use of the cluster population properties to constrain cosmological parameters has been widely used in the past. One fundamental limitation of this approach is that it usually relies on the knowledge of the scaling relations, including their calibration. For instance, the local abundance of clusters is a sensitive probe of the amplitude of matter fluctuations $\sigma_8$. However, it is well known that the amplitude of $\sigma_8$ is degenerated with the normalisation constant $A_{TM}$ \citep{Blanchard2005}. In the present work we follow a different strategy: we perform a self-consistent analysis to constrain  both the cosmological parameters of the $\Lambda CDM$  model and the parameters of the mass-temperature scaling law of X-ray clusters simultaneously. 

To explore the full parameter space and to remove any degeneracy, we need to combine several relevant data sets. In this analysis, we adopted the standard $\Lambda CDM$ cosmology as our reference and performed a MCMC analysis using the COSMOMC package \citep{Lewis2002,Lewis2013}. We combined the constraining power of the CMB data from Planck \citep{Planck2013_I} as well as our measured temperature function through the use of a separate COSMOMC module of our design. 

In practice, we do not use the measured temperature function as it is, i.e. as defined in Eqs.~(\ref{eq:temp_fct1}) to (\ref{eq:temp_fct3}) and shown as red points in Fig.~\ref{fig:n_T_fit_data}. Instead, we binned the temperature range covered by our sample of clusters, and define a binned version of the temperature function  defined as~
\begin{equation}
        n_{binned}(T_1,T_2) = n(>T_1) - n(>T_2) 
\end{equation}
for a given temperature interval $[T_1,T_2]$. This is equivalent to computing the sum of the inverse volumes of all the clusters detected in a given $[T_1,T_2]$ temperature interval 
\begin{equation}
        n_{binned}(T_1,T_2) = \sum_{T_1<T_i<T_2}\frac{1}{V(T_i)} 
.\end{equation}
We used ten distinct temperature bins for the likelihood analysis, defined such that each bin contains approximatively the same amount of clusters from the sample.

In our analysis, we need to estimate the likelihood of any given set of cosmological parameters given the measured temperature function. The details of this likelihood is hard to determine analytically. To circumvent this problem, we perform a bootstrap analysis on our sample of 73 selected clusters, i.e.~ 
\begin{itemize}
        \item we build a ``new" sample by drawing randomly $73\pm\sqrt{73}$ clusters from our original sample;
        \item we recompute the binned temperature function, keeping the same $T$ bins;
and        \item we repeat the procedure a large number of times.
\end{itemize}
Therefore, for each temperature bin we can compute a histogram of all the values of the temperature function coming from the bootstrap process (cf. Fig.~\ref{fig:boot_plot}). The mean value is obviously the measured density of clusters from the original sample in the considered temperature bin. We find that for all bins, the distribution of the values are very well fitted by a Weibull distribution, defined as~
\begin{equation}
        f(x)= \frac{k}{\lambda}\left(\frac{x}{\lambda}\right)^{k-1}e^{-(x/\lambda)^{k}}.
\end{equation}
The fitting is illustrated in Fig.~\ref{fig:boot_plot} for one of our temperature bins. The mean value of random variables following this distribution is related to the shape parameter $k$ and the scale parameter $\lambda$ by~: $<x> = \lambda\, \Gamma(1+1/k)$ where $\Gamma$ is the Gamma function. 

In this study, we assume that the shape $k$ of this distribution (obtained by bootstrapping) is a good approximation to the shape of the likelihood function corresponding to the real, underlying temperature function. Therefore, if a given cosmological model predicts a value of $n_{theo}^i$ for the temperature function in a given temperature bin $i$, then we define the likelihood of this model given the measured value $n_{obs}^i$ as~
\begin{align}
        L(n_{theo}^i|n_{obs}^i) &= P(n_{obs}^i|n_{theo}^i) \\
        &= \frac{k_i}{\lambda_i}\left(\frac{n_{obs}^i}{\lambda_i}\right)^{k_i-1}e^{-(n_{obs}^i/\lambda_i)^{k_i}},
\end{align}
where $k_i$ was determined through the bootstrap process and $\lambda_i=n_{theo}^i/\Gamma(1+1/k_i)$. Finally, the total likelihood for all temperature bins can simply be written as the product of the individual likelihoods for each bin. 

\begin{figure}[ht!]
\centering
\includegraphics[width=\columnwidth]{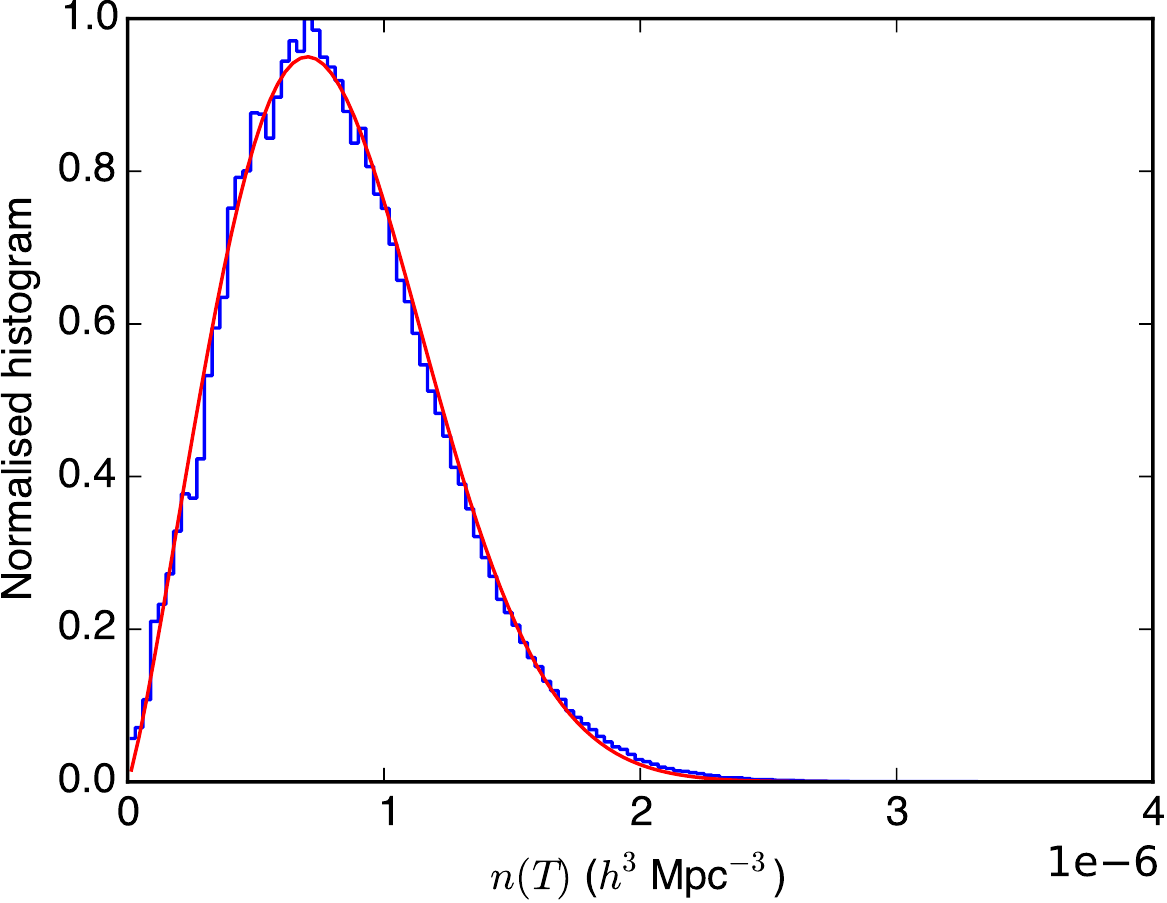}
\caption{Normalised histogram of the values of the temperature function (blue curve), in one of our ten temperature bins, as derived from our bootstrap process (see text for details). The histogram is very well fitted by a Weibull distribution (red curve).}
\label{fig:boot_plot}
\end{figure}

%%%%%%%%%%%%%%%%%%%%%%%%%%%%%%%%%%%%%%%%%%%%%%%%%%%%%%%%%%%%%%%%%%%%%%%%%%%%%%%%%%%%%%%%%%%%%%%%%%%%%%%%%%%%%%%%%%%%%%%%%%%%%%%%%%%%%%%%%%%%%%%%%%%%%%%%%%%%%%%%%%%%%%%%%%
%%%%%%%%%%%%%%%%%%%%%%%%%%%%%%%%%%%%%%%%%%%%%%%%%%%%%%%%%%%%%%%%%%%%%%%%%%%%%%%%%%%%%%%%%%%%%%%%%%%%%%%%%%%%%%%%%%%%%%%%%%%%%%%%%%%%%%%%%%%%%%%%%%%%%%%%%%%%%%%%%%%%%%%%%%

\subsection{First results}
\label{subsec:results}

\begin{table*}[ht!]
\begin{tabular}{@{}cc@{\hspace{0.1cm}}cc@{\hspace{0.1cm}}cc@{\hspace{0.1cm}}cc@{\hspace{0.1cm}}cc@{\hspace{0.1cm}}c@{}}
\hline 
Parameter & \multicolumn{10}{c}{Planck+WP 2013 data} \\[0.2cm] 
 & \multicolumn{2}{c}{only} & \multicolumn{2}{c}{+X-ray (SMT)} & \multicolumn{2}{c}{+X-ray (Tinker $M_{vir}$)} & \multicolumn{2}{c}{+X-ray (Tinker $M_{500c}$)} & \multicolumn{2}{c}{+X-ray (Tinker $M_{500c}$)} \\
 & & & & & & & & & \multicolumn{2}{c}{+free $\beta_{TM}$} \\
\hline
 & Best-fit & 68\% limits & Best-fit & 68\% limits & Best-fit & 68\% limits & Best-fit & 68\% limits & Best-fit & 68\% limits \\
\hline
 & & & & & & & & & & \\[-0.1cm]
$\Omega_{\Lambda}$ & $0.683$ & $[0.668,0.701]$ & $0.684$ & $[0.668,0.701]$ & $0.684$ & $[0.670,0.703]$ & $0.683$ & $[0.671,0.705]$ & $0.685$ & $[0.668,0.700]$ \\
$\Omega_m$ & $0.317$ & $[0.299,0.332]$ & $0.316$ & $[0.299,0.332]$ & $0.316$ & $[0.297,0.330]$ & $0.317$ & $[0.295,0.329]$ & $0.315$ & $[0.300,0.332]$ \\
$n_s$ & $0.961$ & $[0.952,0.966]$ & $0.961$ & $[0.952,0.967]$ & $0.961$ & $[0.953,0.967]$ & $0.961$ & $[0.953,0.967]$ & $0.962$ & $[0.952,0.967]$ \\
$\sigma_8$ & $0.831$ & $[0.816,0.841]$ & $0.830$ & $[0.815,0.840]$ & $0.832$ & $[0.815,0.840]$ & $0.832$ & $[0.815,0.840]$ & $0.830$ & $[0.815,0.840]$ \\
$H_0$ & $67.1$ & $[66.0,68.3]$ & $67.2$ & $[66.1,68.4]$ & $67.2$ & $[66.1,68.5]$ & $67.1$ & $[66.2,68.6]$ & $67.3$ & $[66.0,68.4]$ \\
$\tau$ & $0.0898$ & $[0.0753,0.102]$ & $0.0889$ & $[0.0752,0.102]$ & $0.0908$ & $[0.0756,0.102]$ & $0.0908$ & $[0.0773,0.103]$ & $0.0899$ & $[0.0753,0.102]$\\[0.1cm]
\hline
 & & & & & & & & & & \\[-0.1cm]
$A_{TM}$ & - & - & $6.88$ & $[6.49,7.44]$ & $7.29$ & $[6.90,7.94]$ & $6.29$ & $[5.97,6.93]$ & $6.71$ & $[6.08,7.83]$ \\
$\beta_{TM}$ & - & - & - & - & - & - & - & - & $0.713$ & $[0.665,0.799]$\\
 & & & & & & & & & & \\[-0.1cm]
\hline
\end{tabular} 
\caption{Statistical analysis of our MCMC runs. For the five scenarios considered in Sect.~\ref{subsec:results}, we give best-fit parameters (i.e. the parameters that maximise the overall likelihood for each data combination) as well as 68\% confidence limits for the considered parameters.}
\label{tab:param_MCMC}
\end{table*}

When used alone, galaxy clusters cannot provide strong constraints on the cosmological parameters because of the degeneracy between $\Omega_m$,  $\sigma_8$, and the calibration of the mass temperature calibration $A_{TM}$. Our approach allows us to infer these parameters in a self-consistent way in a given cosmological paradigm (the flat $\Lambda CDM$ model in our case). 

To do this, we combine the data from cosmological probes (Planck CMB data) and our estimation of the local temperature function of X-ray clusters, using the COSMOMC package. We performed our MCMC analysis in the following five different cases:
\begin{itemize}
        \item with the constraints from CMB data only (thus the mass temperature calibration is not evaluated);
        \item with both CMB and clusters constraints, using the SMT mass function, which uses the virial mass of clusters;
        \item with both CMB and clusters constraints, using the \citet{Tinker2008} mass function, which uses the virial masses when defining clusters;
        \item with both CMB and clusters constraints, using the \citet{Tinker2008} mass function, which uses the critical $M_{500}$ masses, and is our fiducial case; and, finally,
        \item the same as above, with an additional degree of freedom,  leaving the index of the $M$-$T$ scaling law as a free parameter, i.e. Eq.~(\ref{eq:MTrelation}) becomes~
        \begin{equation}
        \label{eq:newMTrelation}
        T=A_{TM}(hM_v)^{\beta_{TM}}\left(\frac{\Omega_{m} \Delta(\Omega_m,z)}{178}\right)^{1/3}(1+z)
        ,\end{equation} with $\beta_{TM}$ as a free parameter.
\end{itemize} 

We chose to use both the SMT and Tinker mass function to quantify the difference in the resulting estimated parameters. Indeed, as mentioned earlier (and shown in Fig.~\ref{fig:mass_fcts_comp}), for the same cosmological model the Tinker function predicts fewer objects than SMT's, which  have an effect on the best-fit parameters found by the MCMC analysis.

Figs.~\ref{fig:clusters_tri_plot} and \ref{fig:clusters_tri_plot2} presents a summary of our MCMC analysis in the form of 1D likelihoods and 2D likelihood contours between the cosmological parameters $\Omega_m$ and $\sigma_8$, the calibration parameter $A_{TM}$, and the scaling index $\beta_{TM}$. The best-fit values and 68\% confidence intervals of these parameters are summarised in Table~\ref{tab:param_MCMC} for all five scenarios.
As we can see, for all of the cases cosmological parameters are nearly identical since their strongest constraints come from the standard CMB data (and not from the cluster abundance). The limits on the values of $A_{TM}$ reflect the uncertainty in the mass function. Its derived value when using either the SMT or the Tinker ($M_{vir}$ version) mass function lies within the $1\sigma$ interval of the other one, with the SMT function producing a lower value of $A_{TM}$. This was expected from the differences in the behaviour of the two mass functions. 
We note that adding an additional degree of freedom with the scaling index $\beta_{TM}$ does not degrade the constraints of $A_{TM}$ (1.5$\times$ larger 68\% interval) too much , while it simultaneously constrains $\beta_{TM}$ fairly well. Our approach is therefore capable of determining with a relatively good accuracy the two main parameters of the $M$-$T$ scaling law.

The comparison of the best-fit theoretical temperature functions with the data (shown in Fig.~\ref{fig:n_T_fit_data}) shows a good agreement at all temperatures, although there seems to be some deficit in the observed abundance of clusters (around 2 keV). The SMT temperature function appears to be a better fit to the data, both at the lowest and highest temperature. However, one has to remember that the fit is performed on the binned observed temperature function: in practice for our ten temperature bins, the $\Delta\chi^2$ between the SMT and Tinker functions is only $\sim0.14$ (in favour of SMT).

%%%%%%%%%%%%%%%%%%%%%%%%%%%%%%%%%%%%%%%%%%%%%%%%%%%%%%%%%%%%%%%%%%%%%%%%%%%%%%%%%%%%%%%%%%%%%%%%%%%%%%%%%%%%%%%%%%%%%%%%%%%%%%%%%%%%%%%%%%%%%%%%%%%%%%%%%%%%%%%%%%%%%%%%%%
%%%%%%%%%%%%%%%%%%%%%%%%%%%%%%%%%%%%%%%%%%%%%%%%%%%%%%%%%%%%%%%%%%%%%%%%%%%%%%%%%%%%%%%%%%%%%%%%%%%%%%%%%%%%%%%%%%%%%%%%%%%%%%%%%%%%%%%%%%%%%%%%%%%%%%%%%%%%%%%%%%%%%%%%%%

\subsection{Implication for the Planck SZ clusters results}
\label{subsec:planckcalib}

\begin{figure}[ht!]
\centering
\includegraphics[width=\columnwidth]{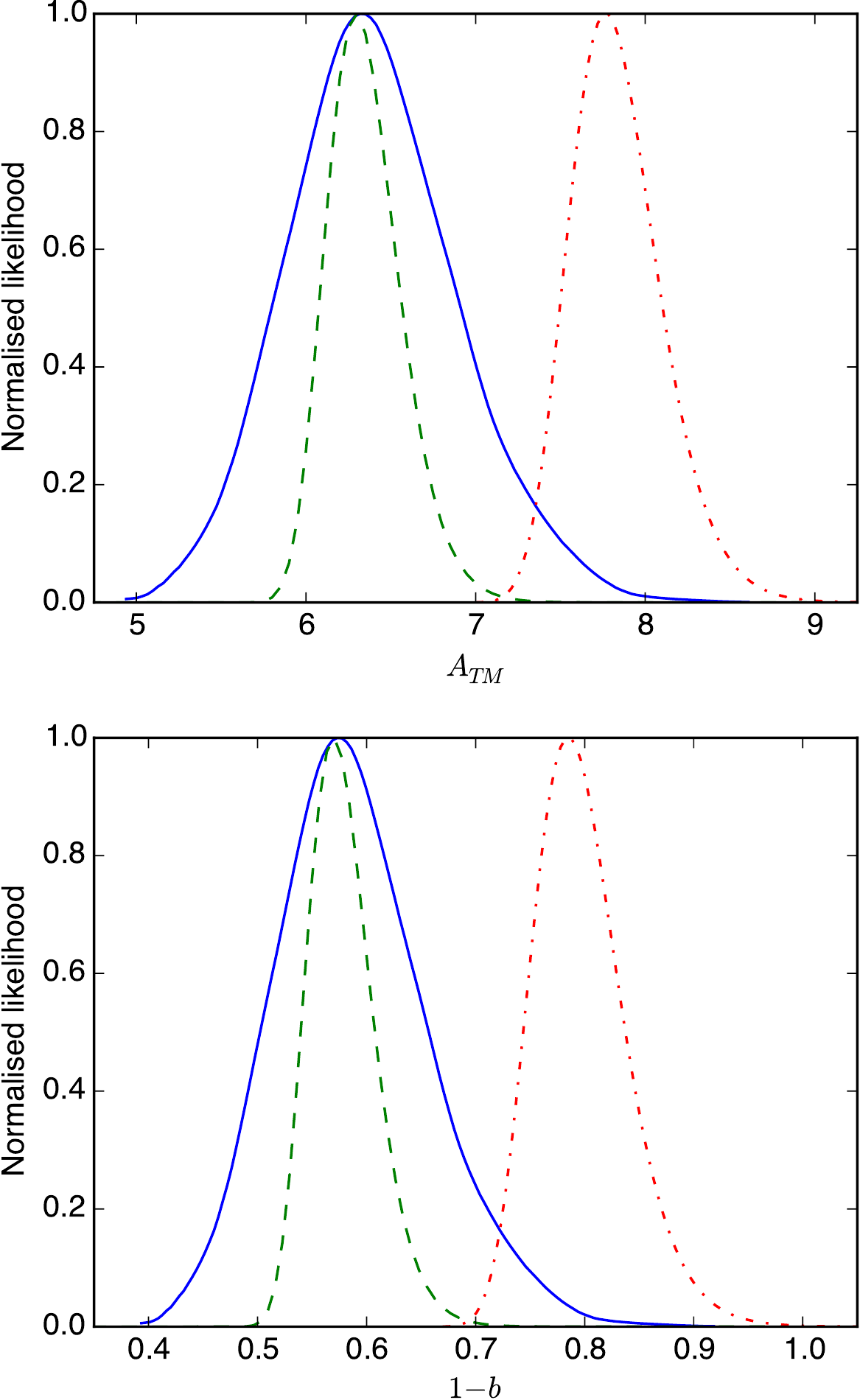}
\caption{\emph{Top panel}: in blue, the 1D likelihood of the $A_{TM}$ parameter derived from our MCMC analysis in our fiducial case (Tinker mass function, with $M_{500c}$ as a definition for the clusters masses. The dashed green and dot-dashed red curves correspond to the 1D likelihood of $A_{TM}$ when fixing all other cosmological and nuisance parameters  to the \emph{Planck CMB} and the \emph{Planck SZ clusters} best-fit cosmology, respectively. \emph{Bottom panel}: corresponding 1D likelihoods of the mass bias $(1-b)$ deduced from the $A_{TM}$ likelihood (see text for details).}
\label{fig:likes}
\end{figure}

\begin{figure}[ht!]
\centering
\includegraphics[width=\columnwidth]{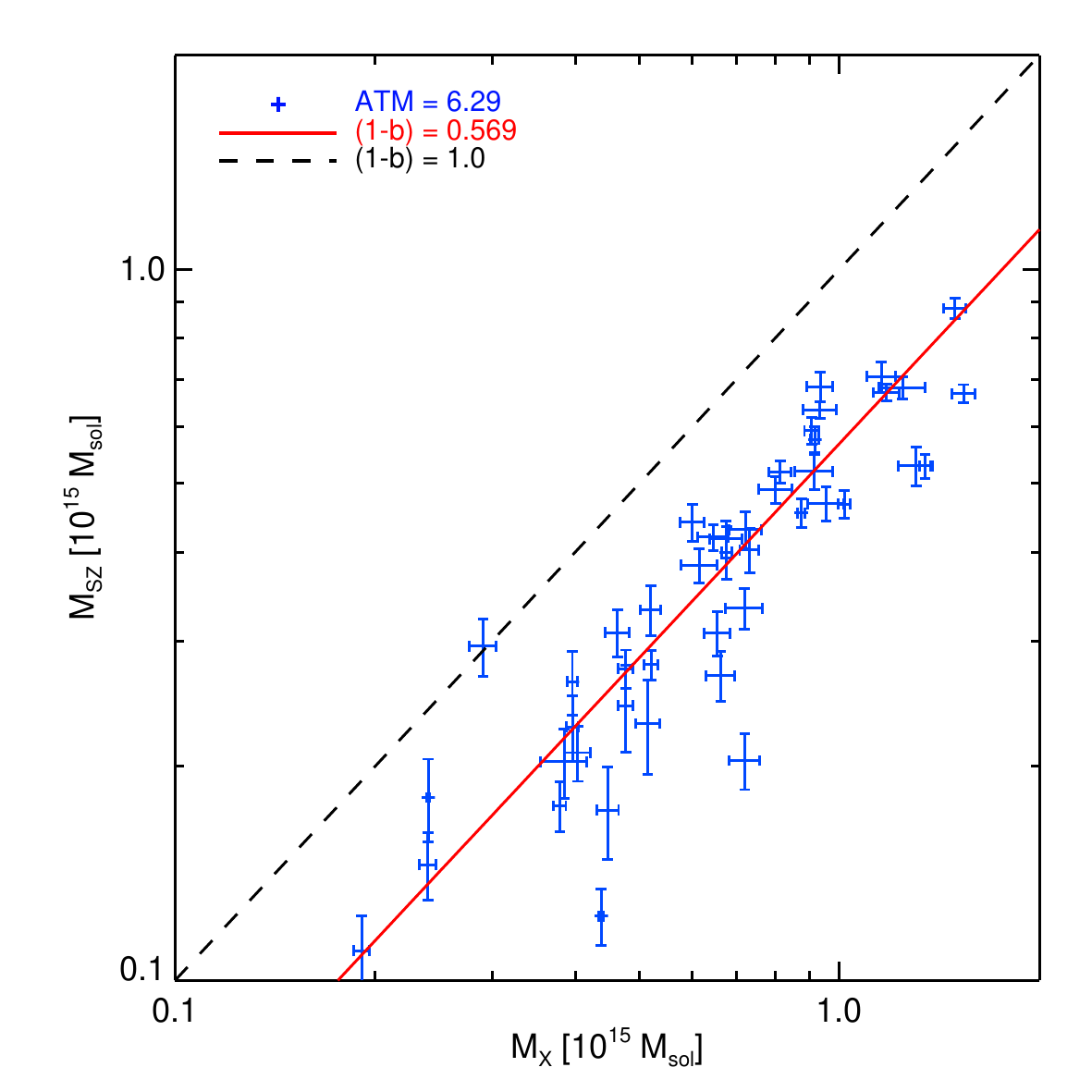}
\caption{For 45 clusters in common between our sample and the Planck sample (see Sec.~\ref{fig:explanation} for details), the blue crosses show the mass estimates from \citetalias{Planck2013_XX} as a function of the masses we derive from their X-ray temperature and our $T_X$-$M$ scaling relation in our fiducial case, with $A_{TM}=6.29$ (best-fit value). The red line corresponds to the best fit from the BCES regression technique, and its slope gives us the mass bias of the Planck masses (see text for the detail).}
\label{fig:explanation}
\end{figure}

In the present section, we now check for consistency between our local sample of X-ray clusters (selected in flux and temperature) and the cluster sample detected through their SZ effect in the Planck CMB data \citepalias{Planck2013_XX}.

The abundance of clusters as a function of redshift is a well-known probe and source of constraints for cosmological models. Although apparently inconsistent results have been obtained from the analyses of (small) X-ray samples \citep{Vauclair2003,2009ApJ...692.1060V}, the use of clusters detected thanks to their SZ signal was anticipated to
be an efficient and reliable method \citep{Barbosa96}. The reasons are twofold: first, the SZ signal does not decline as rapidly with redshift as the X-ray flux, and, second, it is more directly sensitive to the total mass of the clusters, whereas the X-ray flux is biased towards central regions \citep{daS04}.

The Planck collaboration \citepalias{Planck2013_XX} has produced a cosmological SZ cluster sample selected on a signal-to-noise ratio basis, in the redshift range $z\in[0$,$1]$. For their cosmological analysis, they build a scaling relation between the SZ flux $Y$ and the mass $M$ from X-ray observations and simulations. The slope of the scaling relation was obtained from the comparison, for a subsample of 71 clusters, of the SZ flux with the estimation of the mass derived from the hydrostatic equilibrium obtained from XMM-Newton observations \citep{PlanckIR2012_III, Planck2013_XXIX}. The normalisation of the $Y$-$M$ relation was calibrated using simulations \citepalias{Planck2013_XX}. The uncertainty in the determination of this normalisation led to the introduction of a mass bias factor $(1-b)$ in order to parametrise the scaling such that $Y \propto A [(1-b)M]^{\beta}$. The mean value as determined by the aforementioned simulations gives a mass bias of $(1-b)=0.8$ with an assumed constant probability in the range $[0.7-1]$. The cosmological parameters, then derived from the Planck SZ clusters, are shown to be in tension with the CMB-only best-fit cosmology and depends highly on the assumption made on the mass bias. To recover the CMB parameters, it was shown by the Planck collaboration \citepalias[][and again recently in \citealt{PXXIV_2015}]{Planck2013_XX} that a value of $(1-b)\sim0.6$ is required for the mass bias, showing the degeneracy between the cosmological parameter (mainly the amplitude of fluctuations, $\sigma_8$) and the normalisation (or bias) of the $Y$-$M$ relation.

For each Planck cluster with redshift, \citet{Planck2013_XXIX} also provides a \emph{SZ \textup{mass estimate}} calibrated on the $Y$-$M$ scaling relation, and for which no bias is assumed (i.e. $b=0$). In our analysis, we were able to determine the $T_X$-$M$ scaling relation for a Planck CMB cosmology : this allows us to compute a \emph{$T_X$ \textup{based   mass}} for each Planck cluster with a measured X-ray temperature. We then compare these masses to the Planck estimates for a subsample corresponding to the common clusters between our reference sample of 73 clusters and the 71 used by Planck to derive the scaling relation\footnote{available at \url{szcluster-db.ias.u-psud.fr}} (amounting for 45 clusters, shown in Fig.~\ref{fig:explanation}). By fitting the relation between the two set of masses \citep[using the BCES regression technique, see][]{akr96}, we deduce the mass bias $(1-b)$. In practice, we did not obtain a single value of our $A_{TM}$ normalisation parameter, but a likelihood function over a range of $A_{TM}$ values. For any value of the $A_{TM}$ normalisation parameter, we can derive the associated mass bias $(1-b)$ : we can therefore translate our $A_{TM}$ likelihood into a $(1-b)$ likelihood (see Fig.~\ref{fig:likes}). Figure \ref{fig:explanation} shows an example for the fitting of the relation between mass estimates, for our best-fit value of $A_{TM}=6.29$ (in our fiducial case) leading to a value of $(1-b)=0.569$. This value is in very good concordance with the value derived by \citepalias{Planck2013_XX} when fitting Planck CMB and the SZ cosmological sample ($1-b=0.6$), thus showing good, indirect agreement between our sample (based on X-ray measurements of local clusters) and the Planck sample (selected by their SZ signal between $z\in[0$,$1]$).

%%%%%%%%%%%%%%%%%%%%%%%%%%%%%%%%%%%%%%%%%%%%%%%%%%%%%%%%%%%%%%%%%%%%%%%%%%%%%%%%%%%%%%%%%%%%%%%%%%%%%%%%%%%%%%%%%%%%%%%%%%%%%%%%%%%%%%%%%%%%%%%%%%%%%%%%%%%%%%%%%%%%%%%%%%
%%%%%%%%%%%%%%%%%%%%%%%%%%%%%%%%%%%%%%%%%%%%%%%%%%%%%%%%%%%%%%%%%%%%%%%%%%%%%%%%%%%%%%%%%%%%%%%%%%%%%%%%%%%%%%%%%%%%%%%%%%%%%%%%%%%%%%%%%%%%%%%%%%%%%%%%%%%%%%%%%%%%%%%%%%
%%%%%%%%%%%%%%%%%%%%%%%%%%%%%%%%%%%%%%%%%%%%%%%%%%%%%%%%%%%%%%%%%%%%%%%%%%%%%%%%%%%%%%%%%%%%%%%%%%%%%%%%%%%%%%%%%%%%%%%%%%%%%%%%%%%%%%%%%%%%%%%%%%%%%%%%%%%%%%%%%%%%%%%%%%

\section*{Conclusion}

The abundance of clusters found in Planck through their SZ signal has been identified as being somewhat surprisingly low. One possible origin of this ``tension'', identified in the original dedicated Planck paper, is the calibration of the mass SZ signal used in the analysis (no tension subsists if the calibration is left free). Indeed, cluster masses were estimated from the use of the hydrostatic equation based on X-ray observations, the so-called hydrostatic masses.  The accuracy of these estimates have been the subject of debate in the past for several reasons: \\ \emph{(i)} hydrostatic solutions may allow for a wide range of masses for a given set of observations and {\em \textup{their uncertainties}} \citep{bb97}. Differences in X-ray instrument calibration and analysis pipelines might lead to additionnal uncertainties \citep{2015A&A...575A..30S}. \\ \emph{(ii)} gas in clusters might not be in hydrostatic equilibrium due for instance to residual turbulent motions of the gas \citep{bn98,nag07,pif08,mene10} \\ \emph{(iii)} the solution to the overcooling problem \citep{Blanchard1992} most likely needs feedback physics, which are still very uncertain and translate into unknown uncertainties in mass estimations derived from X-ray data. \\ More recently, comparisons made to masses obtained from lensing confirmed that hydrostatic masses seem to yield underestimated values, but no real convergence on lensing masses has been achieved (see \citealt{2014MNRAS.439...48A} versus \citealt{2015MNRAS.449..685H} and \citealt{2015arXiv150704385U}). Yet, the use of a larger sample of clusters detected by Planck \citep{PXXIV_2015} leads to essentially the same tension. The subject of masses, dispersion, and biases in cluster masses estimation is currently an actively discussed subject  \citep{2014MNRAS.438...78R,Se15}.

\medskip

In this paper, we have explored the constraints that can be set on clusters masses in $\Lambda$CDM cosmology matching the Planck CMB data. Our approach was to use the local temperature distribution function of X-ray clusters, and to constrain the mass calibration of scaling laws by requiring that inferred theoretical abundance of local clusters are consistent with observations. This approach is essentially independent of that followed in Planck clusters analyses, and therefore offers an interesting cross-check. Our approach allows us to determine the calibration of the mass-temperature relation and its uncertainty, in $\Lambda$CDM cosmology, in a way that is fully consistent with CMB data. The addition of further cosmological constraints such as BAO of SNIa Hubble diagram would not significantly modify these figures, as they do not provide additional constraints on matter fluctuations. Using the \citet{Tinker2008} mass function, we derived the calibration of the mass-temperature at the virial radius as well as at the standard $R_{500c}$ radius. Our results are summarized in Table~\ref{tab:param_MCMC}. As we tested the $M$-$T$ scaling law, not only were we able to constrain its normalisation $A_{TM}$ but also its slope $\beta_{TM}$, highlighting the constraining power of our approach and the consistency of the scaling law. It should be noted that during the refereeing process of the present article, the 2015 Planck CMB likelihood was publicly released. The resulting new constraints on the scaling law parameters are  tighter, but only by a small margin (as illustrated in Fig.~\ref{fig:clusters_tri_plot3} of the appendix), and do not change our conclusions presented here.

\medskip

The main findings of our analysis is that masses of clusters are 75\% higher (mass here referring to the $M_{500c}$ commonly used in scaling relations), translating in the so-called mass bias $1-b$ = 0.569. Since this bias has been obtained from an analysis based on X-ray counts, and is thus independent of the Planck SZ counts, it is an important test of the consistency on the calibration needed in $\Lambda$CDM to fit the observed abundance of clusters. Indeed, our calibration based on the abundance of X-rays clusters is essentially consistent with that inferred from SZ counts in Planck. This leaves us with the conclusion that cluster counts need a significant revision of one the ingredients of the standard modelling of clusters in $\Lambda$CDM : the physics of clusters might be more complex than previously assumed and their mass have been systematically underestimated, or the mass function used -- based only on dark matter simulations -- significantly overestimates cluster abundance. A last possibility is that there is indeed a more serious problem in the cosmological model and the amplitude of the matter power spectrum $\sigma_8$ is actually lower, calling for extensions of the $\Lambda$CDM model. In this respect, accurate observational calibration of the actual mass of clusters appears as an important clue for solving this issue.

\begin{acknowledgements}
  This work has been carried out thanks to the support of the OCEVU Labex (ANR-11-LABX-0060) and the A*MIDEX project (ANR-11-IDEX-0001-02) funded by the ``Investissements d'Avenir'' French government program managed by the ANR. Authors would like to thank M. Arnaud, E. Pointecouteau, and G. W. Pratt for providing tools for the BCES fit. This research has made use  of the X-Rays Clusters Database (BAX), which is operated at the Institut de Recherche en Astrophysique et Plan\'etologie (IRAP) and of the SZ-Cluster Database operated by the Integrated Data and Operation Center (IDOC) at the Institut d'Astrophysique Spatiale (IAS) under contract with CNES and CNRS.
\end{acknowledgements}

%%%%%%%%%%%%%%%%%%%%%%%%%%%%%%%%%%%%%%%%%%%%%%%%%%%%%%%%%%%%%%%%%%%%%%%%%%%%%%%%%%%%%%%%%%%%%%%%%%%%%%%%%%%%%%%%%%%%%%%%%%%%%%%%%%%%%%%%%%%%%%%%%%%%%%%%%%%%%%%%%%%%%%%%%%
%%%%%%%%%%%%%%%%%%%%%%%%%%%%%%%%%%%%%%%%%%%%%%%%%%%%%%%%%%%%%%%%%%%%%%%%%%%%%%%%%%%%%%%%%%%%%%%%%%%%%%%%%%%%%%%%%%%%%%%%%%%%%%%%%%%%%%%%%%%%%%%%%%%%%%%%%%%%%%%%%%%%%%%%%%
%%%%%%%%%%%%%%%%%%%%%%%%%%%%%%%%%%%%%%%%%%%%%%%%%%%%%%%%%%%%%%%%%%%%%%%%%%%%%%%%%%%%%%%%%%%%%%%%%%%%%%%%%%%%%%%%%%%%%%%%%%%%%%%%%%%%%%%%%%%%%%%%%%%%%%%%%%%%%%%%%%%%%%%%%%

\bibliographystyle{aa}
\bibliography{Bibliography}

%%%%%%%%%%%%%%%%%%%%%%%%%%%%%%%%%%%%%%%%%%%%%%%%%%%%%%%%%%%%%%%%%%%%%%%%%%%%%%%%%%%%%%%%%%%%%%%%%%%%%%%%%%%%%%%%%%%%%%%%%%%%%%%%%%%%%%%%%%%%%%%%%%%%%%%%%%%%%%%%%%%%%%%%%%
%%%%%%%%%%%%%%%%%%%%%%%%%%%%%%%%%%%%%%%%%%%%%%%%%%%%%%%%%%%%%%%%%%%%%%%%%%%%%%%%%%%%%%%%%%%%%%%%%%%%%%%%%%%%%%%%%%%%%%%%%%%%%%%%%%%%%%%%%%%%%%%%%%%%%%%%%%%%%%%%%%%%%%%%%%
%%%%%%%%%%%%%%%%%%%%%%%%%%%%%%%%%%%%%%%%%%%%%%%%%%%%%%%%%%%%%%%%%%%%%%%%%%%%%%%%%%%%%%%%%%%%%%%%%%%%%%%%%%%%%%%%%%%%%%%%%%%%%%%%%%%%%%%%%%%%%%%%%%%%%%%%%%%%%%%%%%%%%%%%%%

\appendix

\section{MCMC likelihood}

In this appendix, we present the results of our MCMC runs for all the scenarios considered in Sect.~\ref{subsec:results}, through ``triangle'' plots  showing the 1D and 2D posterior distributions for the calibration parameter $A_{TM}$, scaling index $\beta_{TM}$, the matter density $\Omega_m$, and the amplitude of the linear power spectrum $\sigma_8$. Fig.~\ref{fig:clusters_tri_plot} illustrates the differences between the various mass functions used throughout our work, while Fig.~\ref{fig:clusters_tri_plot2} shows the impact of having the scaling index $\beta_{TM}$ as a additional free parameter. Fig.~\ref{fig:clusters_tri_plot} presents the impact of the latest Planck 2015 data on our results obtained with the 2013 data.

\begin{figure*}[h]
        \centering
        \includegraphics[width=\textwidth]{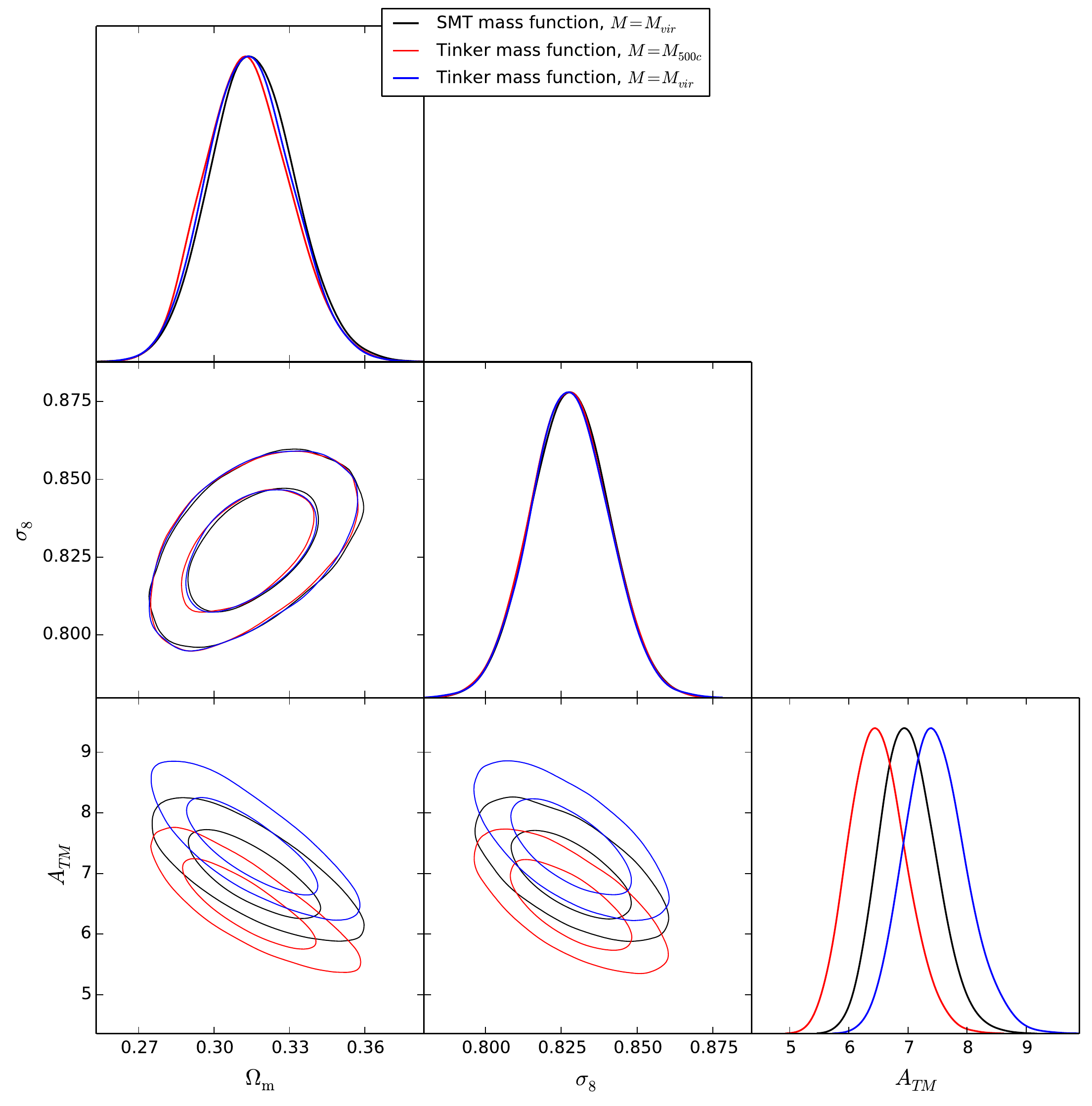}
        \caption{1D posterior distributions and 2D confidence (68\% and 95\%) contours for the parameters $A_{TM}$, $\Omega_m$, and $\sigma_8$, as derived from our MCMC analysis in three different cases: contraints from CMB$+$clusters using either the SMT mass function (red), the Tinker mass function with a virial mass definition (blue) or with $M_{500}$ critical masses (black). The statistical analysis of the corresponding MCMC chains is given in Table~\ref{tab:param_MCMC}.}
        \label{fig:clusters_tri_plot}
\end{figure*}

\begin{figure*}[h]
        \centering
        \includegraphics[width=\textwidth]{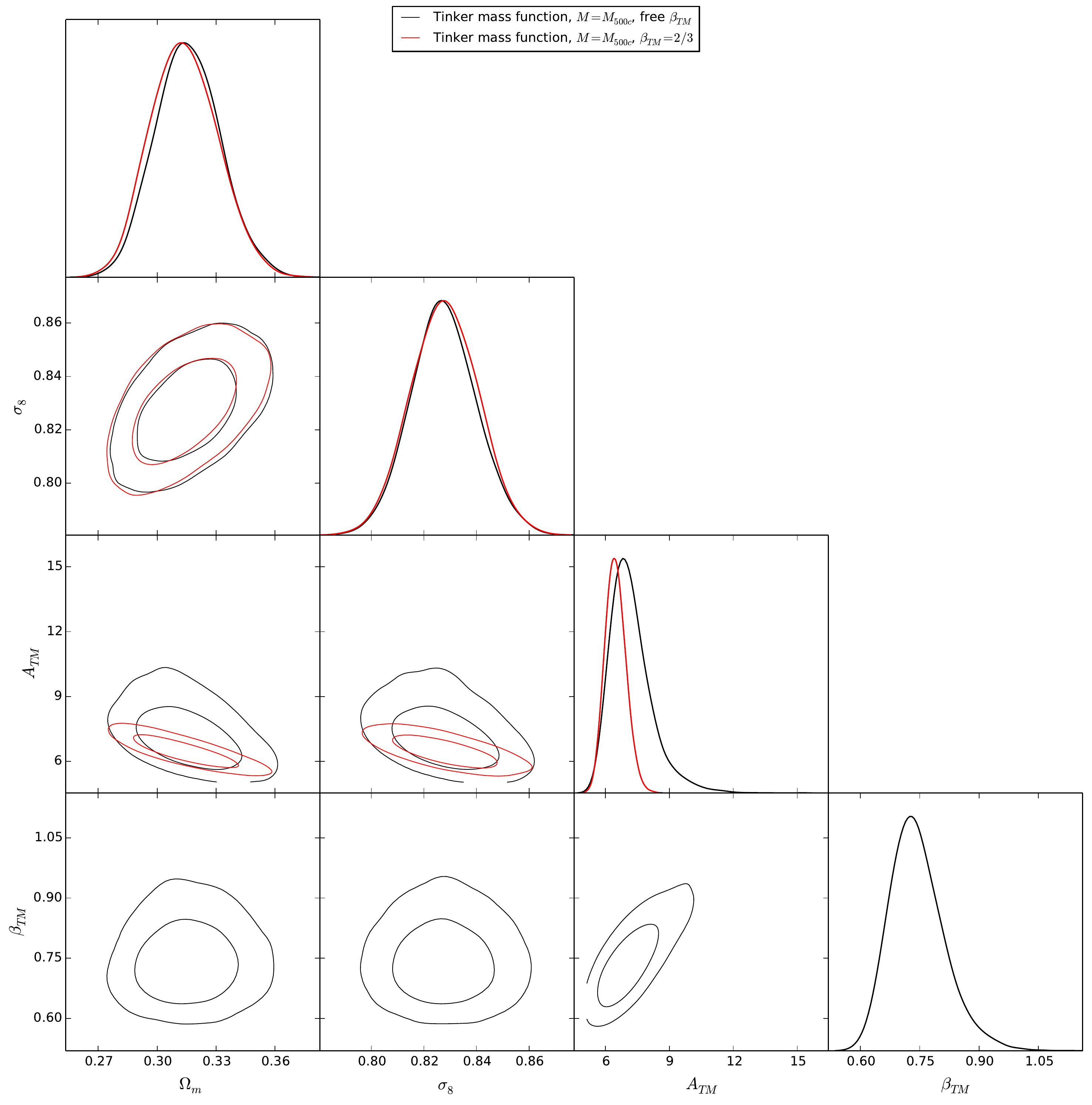}
        \caption{1D posterior distributions and 2D confidence (68\% and 95\%) contours for the parameters $A_{TM}$, $\beta_{TM}$, $\Omega_m$, and $\sigma_8$, as derived from our MCMC analysis in two different cases: contraints from CMB$+$clusters using the Tinker mass function with $M_{500}$ critical masses, with a fixed scaling index $\beta_{TM}=2/3$ (red) or with $\beta_{TM}$ as a free parameter (black). The statistical analysis of the corresponding MCMC chains is given in Table~\ref{tab:param_MCMC}.}
        \label{fig:clusters_tri_plot2}
\end{figure*}

\begin{figure*}[h]
        \centering
        \includegraphics[width=\textwidth]{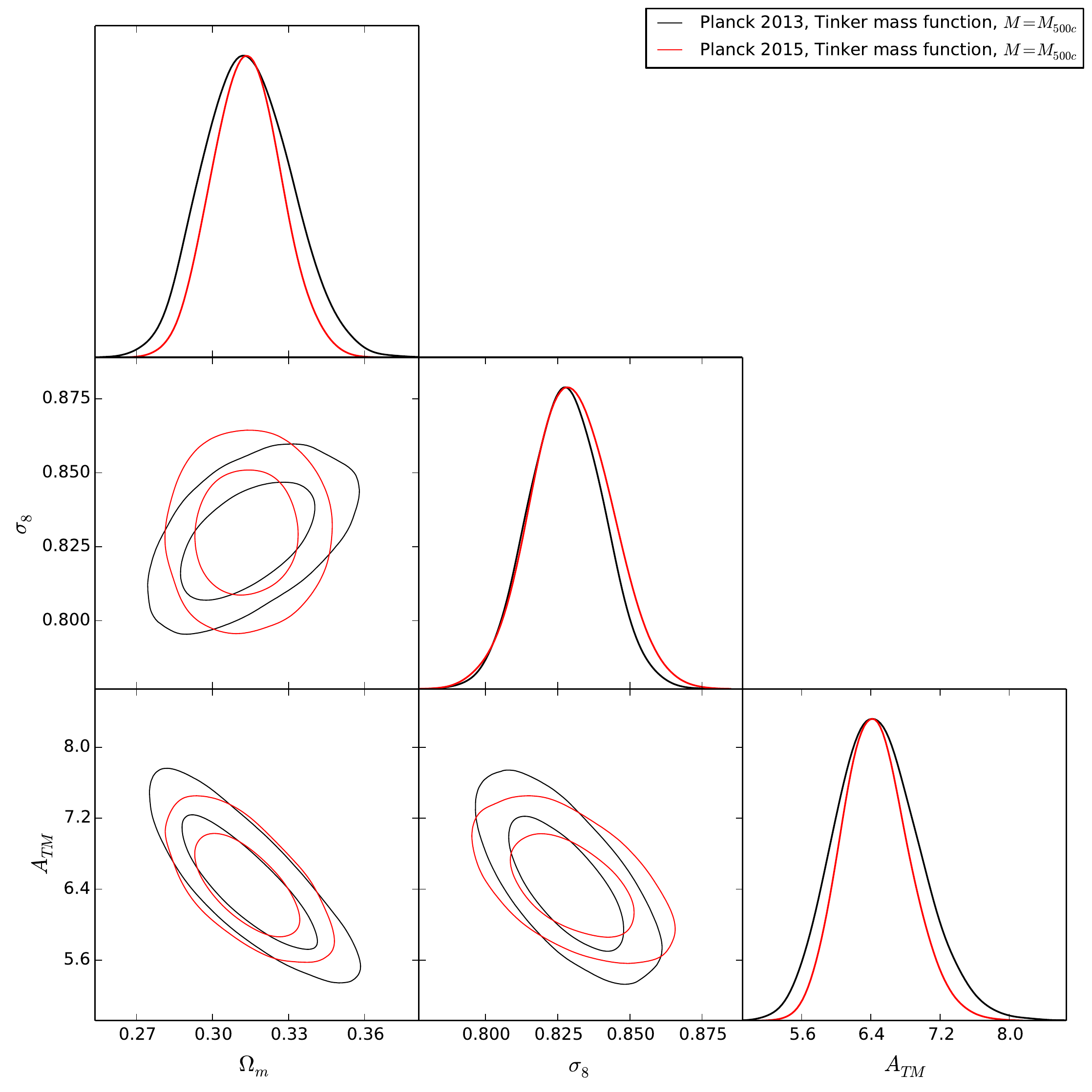}
        \caption{1D posterior distributions and 2D confidence (68\% and 95\%) contours for the parameters $A_{TM}$, $\Omega_m$, and $\sigma_8$, as derived from our MCMC analysis using the Tinker mass function with $M_{500}$ critical masses, with the Planck 2013 (black) and Planck 2015 (red) data sets. The 68\% confidence limits of the scaling law calibration $A_{TM}$ are $[6.04,6.80]$ when using the 2015 CMB data (to be compared to the slightly wider interval $[5.97, 6.93]$ with the 2013 data).}
        \label{fig:clusters_tri_plot3}
\end{figure*}

\end{document}